\documentclass[author-year, 3p]{elsarticle} 
\usepackage[hyphens]{url}
\usepackage{lineno} 
\biboptions{sort&compress} 
\usepackage{graphicx}
\usepackage{booktabs} 
\makeatletter
\def\ps@pprintTitle{%
 \let\@oddhead\@empty
 \let\@evenhead\@empty
 \def\@oddfoot{\it \hfill\today}%
 \let\@evenfoot\@oddfoot}
\makeatother

\usepackage[T1]{fontenc}
\usepackage{lmodern}
\usepackage{amssymb,amsmath}
\usepackage{ifxetex,ifluatex}
\usepackage{fixltx2e} 
\IfFileExists{upquote.sty}{\usepackage{upquote}}{}
\ifnum 0\ifxetex 1\fi\ifluatex 1\fi=0 
  \usepackage[utf8]{inputenc}
\else 
  \usepackage{fontspec}
  \ifxetex
    \usepackage{xltxtra,xunicode}
  \fi
  \defaultfontfeatures{Mapping=tex-text,Scale=MatchLowercase}
  
\fi
\IfFileExists{microtype.sty}{\usepackage{microtype}}{}
\usepackage{longtable}
\usepackage{graphicx}
\makeatletter
\def\maxwidth{\ifdim\Gin@nat@width>\linewidth\linewidth
\else\Gin@nat@width\fi}
\makeatother
\let\Oldincludegraphics\includegraphics
\renewcommand{\includegraphics}[1]{\Oldincludegraphics[width=\maxwidth]{#1}}
\ifxetex
  \usepackage[setpagesize=false, 
              unicode=false, 
              xetex]{hyperref}
\else
  \usepackage[unicode=true]{hyperref}
\fi
\hypersetup{breaklinks=true,
            bookmarks=true,
            pdfauthor={},
            pdftitle={Avoiding tipping points in fisheries management through Gaussian Process Dynamic Programming},
            colorlinks=true,
            urlcolor=blue,
            linkcolor=magenta,
            pdfborder={0 0 0}}
\begin{document}
\begin{frontmatter}

  \title{Avoiding tipping points in fisheries management through Gaussian Process
Dynamic Programming}
    \author[cstar]{Carl Boettiger\corref{c1}}
   \ead{cboettig(at)gmail.com} 
   \cortext[c1]{Corresponding author}
    \author[cstar]{Marc Mangel}

    \author[noaa]{Stephan Munch}

      \address[cstar]{Center for Stock Assessment Research, Department of Applied Math and
Statistics, University of California, Mail Stop SOE-2, Santa Cruz, CA
95064, USA}    
    \address[noaa]{Southwest Fisheries Science Center, National Oceanic and Atmospheric
Administration, 110 Shaffer Road, Santa Cruz, CA 95060, USA}    
  
  \begin{abstract}
  Model uncertainty and limited data are fundamental challenges to robust
  management of human intervention in a natural system. These challenges
  are acutely highlighted by concerns that many ecological systems may
  contain tipping points, such as Allee population sizes. Before a
  collapse, we do not know where the tipping points lie, if they exist at
  all. Hence, we know neither a complete model of the system dynamics nor
  do we have access to data in some large region of state-space where such
  a tipping point might exist. We illustrate how a Bayesian Non-Parametric
  (BNP) approach using a Gaussian Process (GP) prior provides a flexible
  representation of this inherent uncertainty. We embed GPs in a
  Stochastic Dynamic Programming (SDP) framework in order to make robust
  management predictions with both model uncertainty and limited data. We
  use simulations to evaluate this approach as compared with the standard
  approach of using model selection to choose from a set of candidate
  models. We find that model selection erroneously favors models without
  tipping points -- leading to harvest policies that guarantee extinction.
  The GPDP performs nearly as well as the true model and significantly
  outperforms standard approaches. We illustrate this using examples of
  simulated single-species dynamics, where the standard model selection
  approach should be most effective, and find that it still fails to
  account for uncertainty appropriately and leads to population crashes,
  while management based on the GPDP does not, since it does not
  underestimate the uncertainty outside of the observed data.
  \end{abstract}
   \begin{keyword} Bayesian \sep Structural Uncertainty \sep Nonparametric \sep Optimal Control \sep Decision Theory \sep Gaussian Processes \sep Fisheries Management \sep \end{keyword}
 \end{frontmatter}

\section{Introduction}\label{introduction}

Decision making under uncertainty is a ubiquitous challenge in the
management of human intervention in natural resources and conservation.
Decision-theoretic approaches provide a framework to determine the best
sequence of actions in face of uncertainty, but only when that
uncertainty can be meaningfully quantified (Fischer et al. 2009). Over
the last four decades (beginning with Clark (1976), Clark (2009) and
Walters and Hilborn (1978)) dynamic optimization methods, particularly
Stochastic Dynamic Programming (SDP), have become increasingly important
as a means of understanding how to manage human intervention into
natural systems. Simultaneously, there has been increasing recognition
of the importance of multiple steady states or `tipping points'
(Scheffer et al. 2001, 2009, Polasky et al. 2011) in ecological systems.

We develop a novel approach to address these concerns in the context of
fisheries; although the challenges and methods are germane to other
problems of conservation or natural resource exploitation. Economic
value and ecological concern have made marine fisheries the crucible for
much of the founding work on management under uncertainty (Gordon 1954,
Clark 1976, 2009, May et al. 1979, Reed 1979, Ludwig and Walters 1982).

Even if we know the proper deterministic description of the biological
system, there is intrinsic stochasticity in biological dynamics,
measurements, and implementation of policy (\emph{e.g.} Reed 1979, Clark
and Kirkwood 1986, Roughgarden and Smith 1996, Sethi et al. 2005). We
may also lack knowledge about the parameters of the biological dynamics
(parametric uncertainty, \emph{e.g.} Ludwig and Walters 1982, Hilborn
and Mangel 1997, McAllister 1998, Schapaugh and Tyre 2013), or even not
know which model is proper description of the system (structural
uncertainty, \emph{e.g.} Williams 2001, Cressie et al. 2009,
Athanassoglou and Xepapadeas 2012). Of these, the latter is generally
the hardest to quantify. Typical approaches confront the data with a
collection of models, assuming that the true dynamics (or reasonable
approximation) is among the collection and then use model choice or
model averaging to arrive at a conclusion (Williams 2001, Cressie et al.
2009, Athanassoglou and Xepapadeas 2012). Even setting aside other
concerns (see Cressie et al. (2009)), these approaches are unable to
describe uncertainty outside the observed data range.

Structural uncertainty is particularly insidious when we try to predict
outside of the range of observed data (Mangel et al. 2001) because we
are extrapolating into unknown regions. In management applications, this
extrapolation uncertainty is particularly important since (a) management
involves considering actions that may move the system outside the range
of observed behavior, and (b) the decision tools (optimal control
theory, SDP) rely on both reasonable estimates of the expected outcomes
and on the weights given to those outcomes (\emph{e.g.} Weitzman 2013).
Thus characterizing uncertainty is as important as characterizing the
expected outcome.

Tipping points in ecological dynamics (Scheffer et al. 2001, Polasky et
al. 2011) highlight this problem because precise models are not
available and data are limited such as around high stock levels or an
otherwise desirable state. With perfect information, one would know just
how far a system could be pushed before crossing the tipping point, and
management would be simple. But we face imperfect models and limited
data and, with tipping points, even small errors can have very large
consequences, as we shall illustrate later. Because intervention may be
too late once a tipping point has been crossed (but see Hughes et al.
(2013)), management is often concerned with avoiding tipping points
before any data about them are available.

The dual concerns of model uncertainty and incomplete data create a
substantial challenge to existing decision-theoretic approaches
(Brozović and Schlenker 2011). We illustrate how Stochastic Dynamic
Programming (SDP) (Mangel and Clark 1988, Marescot et al. 2013) can be
implemented using a Bayesian Non-Parametric (BNP) model of population
dynamics (Munch et al. 2005a). The BNP method has two distinct
advantages. First, using a BNP model sidesteps the need for an accurate
model-based description of the system dynamics. Second, a BNP model
better reflects uncertainty when extrapolating beyond the observed data.
This is crucial to providing robust decision-making when the correct
model is not known (as is almost always true). {[}We use \emph{robust}
to characterize approaches that provide nearly optimal solutions without
being sensitive to the choice of the (unknown) underlying model.{]}

This paper is the first ecological application of the SDP without an
\emph{a priori} model of the underlying dynamics. Unlike parametric
approaches that can only reflect uncertainty in parameter estimates, the
BNP method provides a broader representation of uncertainty, including
uncertainty beyond the observed data. We will show that Gaussian Process
Dynamic Programming (GPDP) allows us to find robust management solutions
in face of limited data without knowing the correct model structure.

For comparisons, we consider the performance of management based on GPDP
against management policies derived under several alternative parametric
models (Reed 1979, Ludwig and Walters 1982, Mangel and Clark 1988).
Rather than compare models in terms of best fit to data, we compare
model performance in the concrete terms of the decision-maker's
objectives.

\section{Approach and Methods}\label{approach-and-methods}

We first describe the requirements of dynamic optimization for the
management of human intervention in natural resource systems. After that
we describe three parametric models for population dynamics and the
Gaussian Process (GP)\footnote{We abbreviate Gaussian Process as GP,
  which refers to the statistical model we use to approximate the
  population dynamics, and we use the term Gaussian Process Dynamic
  Programming (GPDP), to refer to the \emph{use} of a GP as the
  underlying process model when solving a Dynamic Programming equation.
  Hence we will refer to the models as: GP, Ricker, Allen, etc, and the
  novel method we put forward here as GPDP.} description of population
dynamics. All computer code used here has been embedded in the
manuscript sources (see Xie (2013)), and an implementation of the GPDP
approach is provided as an accompanying R package. Source code, R
package and the CSV data files corresponding each figure are archived in
the supplement (Boettiger et al. 2014).

\subsubsection{Requirements of dynamic
optimization}\label{requirements-of-dynamic-optimization}

Dynamic optimization requires characterizing the dynamics of a state
variable (or variables), a control action, and a value function. For
simplicity, we consider only a single state variable. This is a
best-case scenario for the parametric models because we simulate
underlying dynamics from one of the three parametric models, whereas in
the natural world we never know the ``true'' model. In addition, by
choosing one-dimensional models with just a few parameters, we limit the
chance that poor performance will be due to inability to estimate
parameters accurately, something that becomes a more severe problem for
higher-dimensional parametric models. Finally, the parametric models we
consider are commonly used in modeling stock-recruitment dynamics or to
model sudden transitions between alternative stable states.

We let $X(t)$ denote the size (numbers or biomass) of the focal
population at time $t$ and assume that in the absence of take its
dynamics are:

\begin{equation}
X(t+1) = Z(t) f(X(t), \mathbf{p}) \label{eq1} 
\end{equation}

Where Z(t) is log-normally distributed process stochasticity (Reed 1979)
and $\mathbf{p}$ is a vector of parameters to be estimated from the
data. We describe the three choices for $f(X(t),\mathbf{p})$ in the next
section. The control action is a harvest or take, $h(t)$, measured in
the same units as $X$, at time $t$. Thus, in the presence of take, the
population size on the right hand side of Eqn 1 is replaced by
$S(t) = X(t) - h(t)$.

To construct the value function, we consider a return when $X(t) = x(t)$
and harvest $h(t) = h$ denoted as the reward, $R(x(t), h)$. For example,
if the return is the harvest at time $t$, then
$R(x(t), h(t)) = \min(x(t), h(t))$. We assume that future harvests are
discounted relative to current ones at a constant rate of discount
$\delta$ and ask for the harvest policy that maximizes total discounted
harvest between the current time $t$ and a final time $T$. That is, we
seek to maximize over choices of harvest
$E [ \sum_{t = 0}^{T}  R(X(t), h(t), t) \delta^t]$, where the state
dynamics are given by Eqn 1 and $E$ denotes the expectation over future
population states.

In order to find that policy, we introduce the value function
$V(x(t), t)$ representing the total discounted catch from time $t$
onwards given that $X(t) = x(t)$. This value function satisfies an
equation of SDP (Mangel and Clark 1988, Clark and Mangel 2000, Clark
2009 Mangel 2014),

\begin{equation}
V(x(t), t) = \max_{h(t)} \lbrace R(h(t), x(t)) + \delta \cdot \mathbf{\mathrm{E}}_{X(t+1)} \left[ V(X(t+1), t+1) | x(t), h(t) \right] \rbrace
\end{equation}

where expectation is taken over all possible values of the next state,
$X(t+1)$, and maximized over all possible choices of harvest, $h(t)$.
That is, at time $t$, when population size is $x(t)$ and harvest $h(t)$
is applied, the immediate return is $R(h(t), x(t))$. When the sole
source of uncertainty is the process stochasticity term, $Z$, and thus
the expectation in Eqn 2 is equivalent to taking expectations over
$Z(t)$. That is

\begin{equation}
\mathbf{\mathrm{E}}_{X(t+1)} \left[ V( X(t+1),t+1) | x(t), h(t) \right] = \mathbf{\mathrm{E}}_{Z(t)} \left[V( Z(t) f(x(t) - h(t))|\mathbf{p}), t+1 | x(t), h(t) \right]
\end{equation}

where the population size after the take is $x(t) - h(t)$, which is then
translated into $X(t+1)$ by Eqn 1 (that is, we replace $X(t+1)$ by
$Z(t) f(x(t)-h(t)|\mathbf{p})$).

When the parameters governing the dynamics are also uncertain, we take
the expectation over the posterior distribution for the parameters:

\begin{equation}
\mathbf{\mathrm{E}}_{X(t+1)} \left[ V(X(t+1), t+1) | x(t), h(t) \right] = \mathbf{\mathrm{E}}_{\mathbf{p}|\mathrm{data}} \{ \mathbf{\mathrm{E}}_{Z(t) | \mathbf{p}, \mathrm{data}} \left[ V(Z(t) f(x(t) - h(t)|\mathbf{p}), t)  \right] \}
\end{equation}

When the underlying population dynamics are unknown (the case of
structural uncertainty), the function $f$ itself is uncertain and the
expectation for the next state includes uncertainty in $f$ as well. That
is

\begin{equation}
\mathbf{\mathrm{E}}_{X(t+1)} \left[ V(X(t+1), t+1) | x(t), h(t) \right] = \mathbf{\mathrm{E}}_{\mathbf{p}|\mathrm{data}} \{ \mathbf{\mathrm{E}}_{f, Z(t) | \mathbf{p}, \mathrm{data}} \left[ V( Z(t) f(x(t) - h(t)| \mathbf{p}), t) \right] \}
\end{equation}

We consider the finite time problem with $T=$ 1000, which we solve using
the standard value iteration algorithm (see Mangel and Clark 1988, Clark
and Mangel 2000).

\subsubsection{Parametric Models}\label{parametric-models}

We consider three candidate parametric models for the population
dynamics: The Ricker model, the Allen model (Allen and Tanner 2005), and
the Myers model (Myers et al. 1995), Eqns \eqref{ricker}-\eqref{myers}.
In all three, we let $K$ denote the carrying capacity and $r$ the
maximum per capita growth rate. The Ricker model has two parameters and
the right hand side of Eqn 1 is

\begin{equation}
f(S(t)|r,K) = S(t) e^{r \left(1 - \frac{S(t)}{K} \right) } \label{ricker}
\end{equation}

The Allen model has three parameters

\begin{equation}
f(S(t)|r, K, X_c) = S(t) e^{r \left(1 - \frac{S(t)}{K}\right)\left(S(t) - X_c\right)} \label{allen}
\end{equation}

where $X_c$ denotes the location of the unstable steady state (i.e., the
tipping point).

The Myers model also has three parameters:

\begin{equation}
f(S(t) | r, K, \theta)  = \frac{r S(t)^{\theta}}{1 + \frac{S(t)^\theta}{K}} \label{myers}
\end{equation}

where $\theta = 1$ corresponds to Beverton-Holt dynamics and $\theta$
\textgreater{} 2 leads to Allee effects and multiple stable states.

The Ricker model does not lead to multiple steady states. The Allen
model resembles the Ricker dynamics with an added Allee effect parameter
(Courchamp et al. 2008), below which the population cannot persist. The
Myers model also has three parameters and contains an Allee threshold,
but unlike the Ricker model saturates at high population size. The
multiplicative log-normal stochasticity in Eqn 1 introduces one
additional parameter $\sigma$ that must be estimated.

Because of our interest in management performance in the presence of
tipping points, all of our simulations are based on the Allen model. The
Allen model is thus the state of nature and is expected to provide the
best-case scenario. The Ricker model is a reasonable approximation of
these dynamics far from the Allee threshold (but lacks threshold
dynamics), while the Myers model shares the essential feature of a
threshold but differs in structure from the Allen model. Throughout, we
refer to the ``True'' model when the underlying parameters \emph{are
known without error}, and refer to the ``Allen'' model when these
parameters have been estimated from the sample data.

We consider a period of 40 in which data are obtained to estimate the
parameters or the GP. This is long enough that the estimates do not
depend on the particular realization, and longer times are not likely to
provide substantial improvement. Each of the models is fit to the same
data (Figure 1).

We inferred posterior distributions for the parameters of each model by
Gibbs sampling (Gelman et al. (2003) implemented in R (R Core Team 2013)
using \texttt{jags}, (Su and Masanao Yajima 2013)). We choose uniform
priors for all parameters of the parametric models (See appendix Tables
S1-S3; R code provided). We show one-step-ahead predictions of these
model fits in Figure 1. We tested each chain for Gelman-Rubin
convergence and results were robust to longer runs. For each simulation
we also applied several commonly used model selection criteria (AIC,
BIC, DIC, see Burnham and Anderson (2002)) to identify the best fitting
model.

Additionally, we compute the maximum likelihood estimate (MLE, as we
will refer to this model in the figures) of the parameters for the
(structurally correct) Allen model. Comparing this to using the
posterior distribution of parameters inferred from MCMC for the same
model gives some indication of the importance of this uncertainty in the
dynamic programming.

\subsubsection{The Gaussian Process
model}\label{the-gaussian-process-model}

The core difference for our purpose between the estimated GP and the
estimated parametric models is that the estimated GP model is defined
explicitly in reference to the observed data. As a result, uncertainty
arises in the GP model not only from uncertainty in the parameters, but
is also increases in regions farther from the observed states, such as
low population sizes in the example illustrated here. The estimated
parametric models, by contrast, are completely specified by the
parameters.

The use of GPs to characterize dynamical systems is relatively new
(Kocijan et al. 2005), and was first introduced in the context
ecological modeling and fisheries management in Munch et al. (2005b). GP
models have subsequently been used to test for the presence of Allee
effects (Sugeno and Munch 2013a), estimate the maximum reproductive rate
(Sugeno and Munch 2013b), determine temporal variation in food
availability (Sigourney et al. 2012), and provide a basis for
identifying model-misspecification (Thorson et al. 2014). An accessible
and thorough introduction to the formulation and use of GPs can be found
in Rasmussen and Williams (2006).

A GP is a stochastic process for which any realization consisting of n
points follows a multivariate normal distribution of dimension $n$. To
characterize the GP we need a mean function and a covariance function.
We proceed as follows.

As before, we assume that the data $X(t)$ are observed with process
stochasticity around a mean function $g(X(t))$

\begin{equation}
X(t+1) = g(X(t)) + \varepsilon,
\end{equation}

where $\varepsilon$ are IID normal random variables with zero-mean and
variance $\sigma^2$. Note that we have chosen to assume additive
stochasticity. While we could consider log-normal stochasticity as in
the parametric models, we make this choice to emphasize that the
Gaussian process approach need not have structurally correct
stochasticity to be effective.

In order to make predictions, we update the GP based on the observed set
of transitions. To do so, we collect the time series of observed states
into a vector of ``current'' states,
$\mathbf{X}_{\textrm{obs}} = \{X(1), \dots, X(T-1)\}$ and a vector of
``next'' states $\mathbf{Y}_{\textrm{obs}} = \{X(2),\dots,X(T)\}$ where
$T$ is the time of the final observation. Conditional on these
observations, the predicted next state, $g(X_p)$, for any given
``current'' state, $X_p$ follows a normal distribution with mean $E$ and
variance $C$ determined using the standard rules for conditioning in
multivariate normals, i.e.

\begin{equation}
E = K(X_p, \mathbf{X}_{\textrm{obs}}) \left(K(\mathbf{X}_{\textrm{obs}},\mathbf{X}_{\textrm{obs}}) - \sigma \mathbf{I}_n \right)^{-1} \mathbf{Y}_{\textrm{obs}}
\end{equation}

and

\begin{equation}
C = K(X_p, X_p) - K(X_p, \mathbf{X}_{\textrm{obs}}) \left(K(\mathbf{X}_{\textrm{obs}},\mathbf{X}_{\textrm{obs}}) - \sigma \mathbf{I} \right)^{-1} K(\mathbf{X}_{\textrm{obs}}, X_p)
\end{equation}

Here $\mathbf{I}_n$ is the $n$ by $n$ identity matrix (i.e.~a matrix
with ones down the diagonal and zeros elsewhere) and $K$ is the
`covariance kernel.' The covariance kernel controls how much influence
one observation has on another. In the present application we use the
squared-exponential kernel which, when evaluated over a pair of vectors,
say $\mathbf{x}$ and $\mathbf{y}$, generates a covariance matrix whose
$i,j$th element is given by

\begin{equation}
K_{i,j}(\mathbf{x}, \mathbf{y}) = \exp\left( \frac{ -(x_i - y_j)^2}{2 \ell^2} \right)
\end{equation}

so that $\ell$ gives the characteristic length-scale over which
correlation between two observations decays. See Rasmussen and Williams
(2006) for other choices of covariance kernels and their properties.
Note that this simple formulation assumes a prior mean of zero. For the
parameters we use inverse Gamma priors on both the length-scale $\ell$
and $\sigma$, thus for example

\begin{equation}
f(\ell; \alpha, \beta) = \frac{\beta^\alpha}{\Gamma(\alpha)} \ell^{-\alpha - 1}\exp\left(-\frac{\beta}{\ell}\right)
\end{equation}

For the prior on $\ell$, $\alpha =$ 10 and $\beta =$ 10. The prior on
$\sigma$, $\alpha =$ 5 and $\beta =$ 5.

We use a Metropolis-Hastings Markov Chain Monte Carlo (Gelman et al.
(2003)) to infer posterior distributions of the parameters of the GP
(Figure S4, code in appendix). Since the posterior distributions differ
substantially from the priors (Figure S4), most of the information in
the posterior comes from the data rather than the prior belief.

\subsection{The method of Gaussian Process Dynamic Programming
(GPDP)}\label{the-method-of-gaussian-process-dynamic-programming-gpdp}

We derive the harvest policy from the estimated GP by inserting it into
a SDP algorithm. Given the GP posteriors, we construct the transition
matrix representing the probability of going to each state $X(t+1)$
given any current state $x(t)$ and any harvest $h(t)$ (See the function
\texttt{gp\_transition\_matrix()} in the provided R package). Given this
transition matrix, we use the same value iteration algorithm as in the
parametric case to determine the optimal policy. In doing so, the
uncertainty in the future state under the GP, $X(t+1)$, includes both
process uncertainty (based on the estimation of $\sigma$) and structural
uncertainty of the posterior collection of curves.

\section{Results}\label{results}

\subsection{Parametric and GP models for population
dynamics}\label{parametric-and-gp-models-for-population-dynamics}

To ensure our results are robust to the choice of parameters, we will
consider 96 different scenarios. To help better understand the process,
we first describe in detail the results of a single scenario.

\begin{figure}[htbp]
\centering
\includegraphics{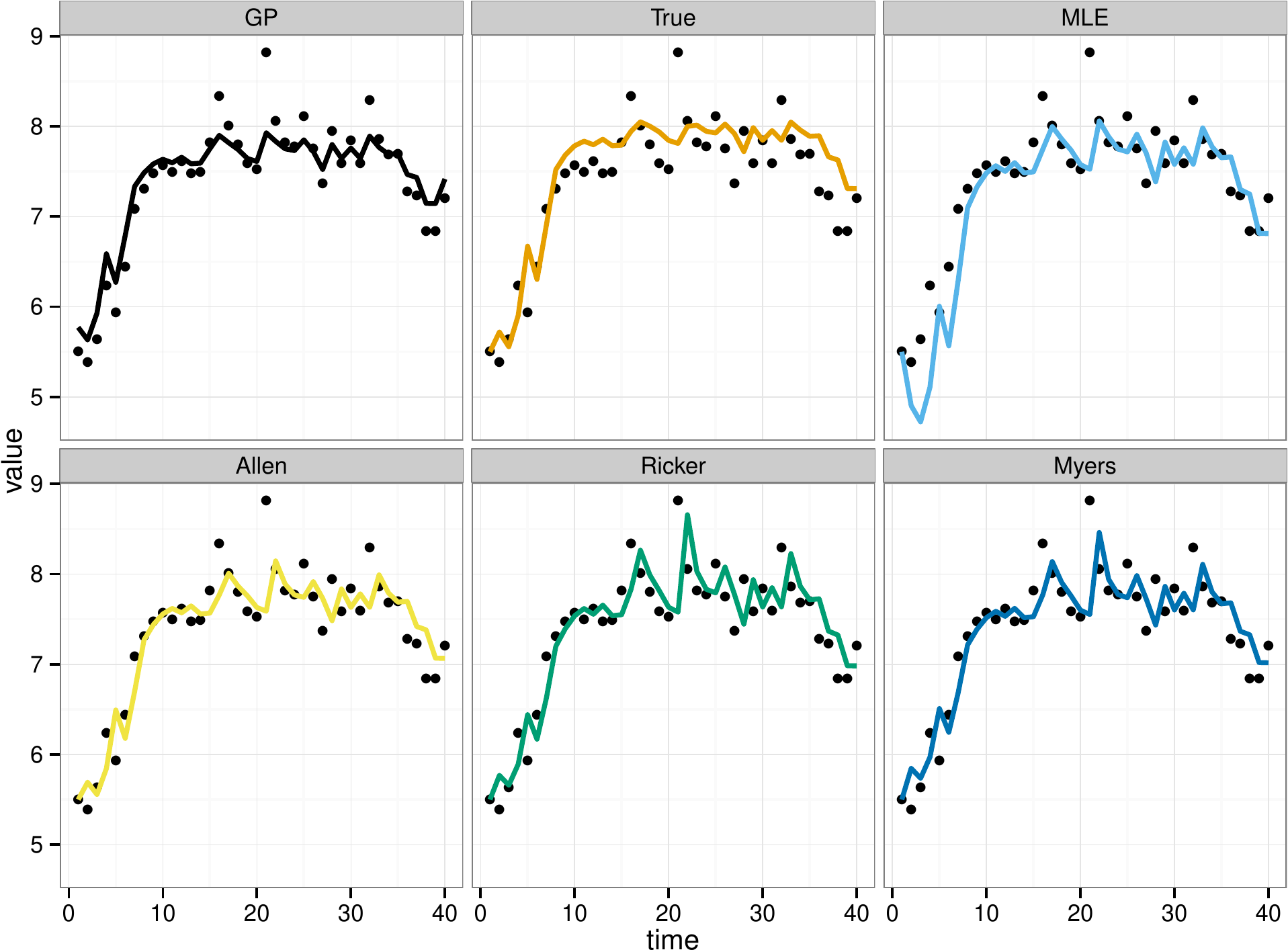}
\caption{Points show the training data of stock-size over time. Curves
show the expectations over the posterior step-ahead predictions based on
each of the estimated models. Observe that all models are fitting the
data reasonably well.)}
\end{figure}

All of the models fit the observed data rather closely and with
relatively small uncertainty. In Figure 1, we show the posterior
predictive curves. The training data of stock sizes observed over time
are points, overlaid with the step-ahead predictions of each estimated
model using the parameters sampled from their posterior distributions.
Compared to the true model most estimates appear to over-fit, predicting
patterns that are actually due purely to stochasticity. Model selection
criteria (Table 1) penalize more complex models and show a preference
for the simpler Ricker model over the models with alternative stable
states (Allen and Myers). Supplement provides details on the model
estimates.

\begin{longtable}[c]{@{}cccc@{}}
\toprule\addlinespace
\begin{minipage}[b]{0.12\columnwidth}\centering
~
\end{minipage} & \begin{minipage}[b]{0.10\columnwidth}\centering
Allen
\end{minipage} & \begin{minipage}[b]{0.11\columnwidth}\centering
Ricker
\end{minipage} & \begin{minipage}[b]{0.11\columnwidth}\centering
Myers
\end{minipage}
\\\addlinespace
\midrule\endhead
\begin{minipage}[t]{0.12\columnwidth}\centering
\textbf{DIC}
\end{minipage} & \begin{minipage}[t]{0.10\columnwidth}\centering
50.75
\end{minipage} & \begin{minipage}[t]{0.11\columnwidth}\centering
50.45
\end{minipage} & \begin{minipage}[t]{0.11\columnwidth}\centering
50.41
\end{minipage}
\\\addlinespace
\begin{minipage}[t]{0.12\columnwidth}\centering
\textbf{AIC}
\end{minipage} & \begin{minipage}[t]{0.10\columnwidth}\centering
-24.51
\end{minipage} & \begin{minipage}[t]{0.11\columnwidth}\centering
-30.13
\end{minipage} & \begin{minipage}[t]{0.11\columnwidth}\centering
-27.01
\end{minipage}
\\\addlinespace
\begin{minipage}[t]{0.12\columnwidth}\centering
\textbf{BIC}
\end{minipage} & \begin{minipage}[t]{0.10\columnwidth}\centering
-17.75
\end{minipage} & \begin{minipage}[t]{0.11\columnwidth}\centering
-25.06
\end{minipage} & \begin{minipage}[t]{0.11\columnwidth}\centering
-20.25
\end{minipage}
\\\addlinespace
\bottomrule
\addlinespace
\caption{Model selection scores for several common criteria (DIC:
Deviance Information Criterion, AIC: Akaike Information Criterion, BIC:
Bayesian Information Criterion) all select the wrong model. As the true
(Allen) model is not distinguishable from the simpler (Ricker) model in
the region of the observed data, this error cannot be avoided regardless
of the model choice criterion. This highlights the danger of model
choice when the selected model will be used outside of the observed
range of the data.}
\end{longtable}

\begin{figure}[htbp]
\centering
\includegraphics{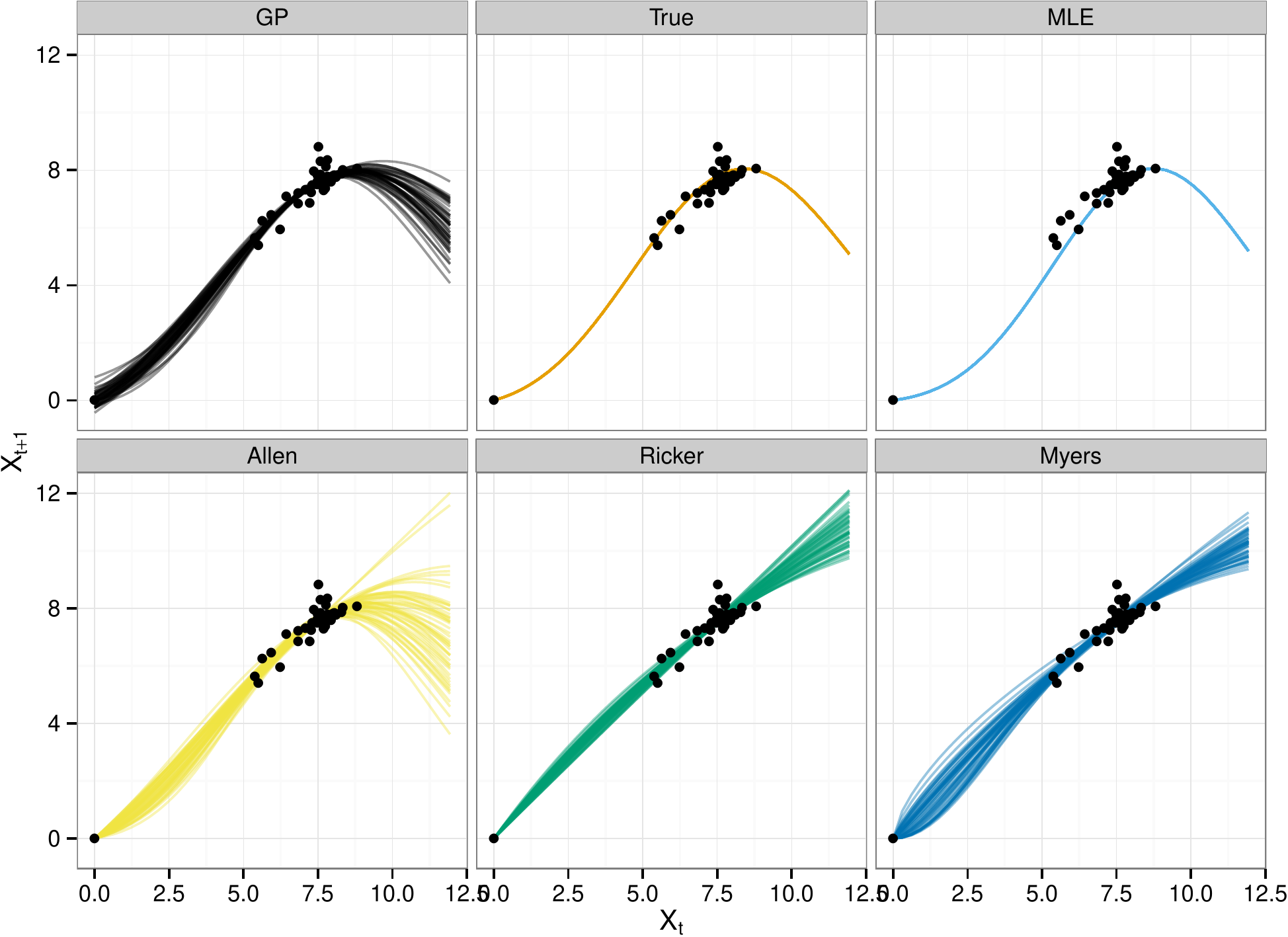}
\caption{The inferred Gaussian process compared to the true process and
maximum-likelihood estimated (MLE) process. We show the expected value
for the function $f$ under each model. Plots show replicates drawn from
the posterior distributions in order to convey uncertainty of the
estimates. Note the MLE is a point estimate of parameters and so
reflects no uncertainty in the distribution. The training data are also
shown as black points. The GP is conditioned on (0,0), shown as a
pseudo-data point.}
\end{figure}

We show the mean inferred population dynamics of each model relative to
the true model used to generate the data in Figure 2, predicting the
relationship between observed population size (x-axis) to the population
size after recruitment the following year. In addition to the raw data,
the GP is conditioned on going through the point 0,0 without error. All
parametric models also make this assumption. Conditioning on (0,0) is
equivalent to making the assumption that the population is closed, so
that once it hits 0 it stays at 0, despite the lack of any data in the
observed sequence to justify this. This assumption illustrates how the
GP can capture common-sense biology without having to assume more
explicit details about the dynamics at low population numbers that have
never been observed. If the population were not closed, one could repeat
the entire analysis without this assumption. Unlike parametric models,
the GP corresponds to a distribution of curves, of which this plot only
shows the means. Uncertainty in the parameters of the GP (not shown)
further widens the band of possible population sizes. In Figure S1 (see
supplement), we show the performance of the models outside the observed
training data.

\begin{figure}[htbp]
\centering
\includegraphics{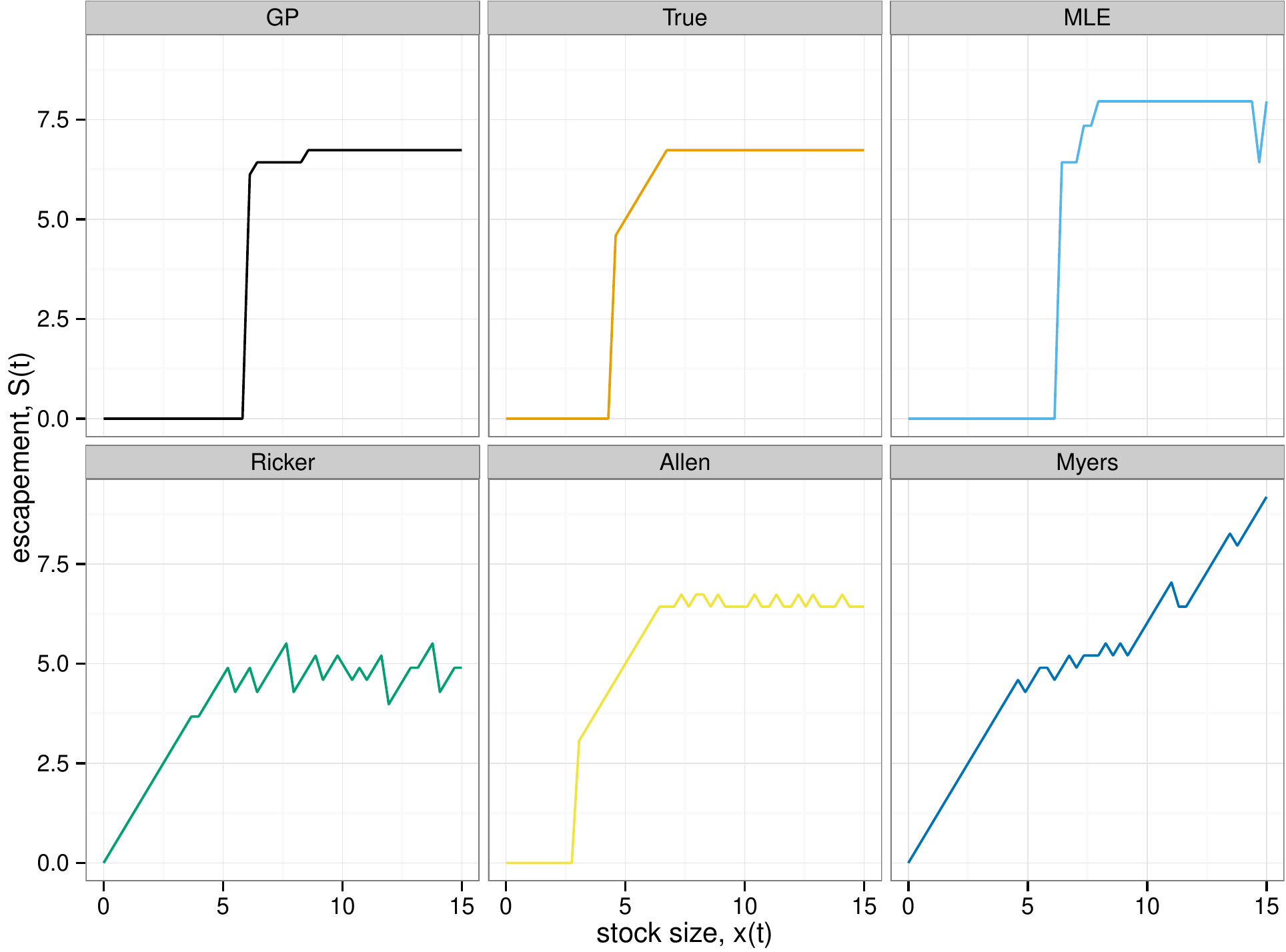}
\caption{The steady-state optimal policy (infinite boundary) calculated
under each model. Policies are shown in terms of target escapement,
$S(t)$, as under models such as this a constant escapement policy is
expected to be optimal (Reed 1979) Several policies show a numerical
jitter due to the discretization of states in the dynamic programming
algorithm; doubling the number of grid points did not qualitatively
change the results.}
\end{figure}

Despite the similarities in model fits to the observed data, the
policies inferred under each model differ widely (Figure 3). Policies
are shown in terms of target escapement, $S(t) = x_t - h$. Under models
such as this a constant escapement policy is expected to be optimal
(Reed 1979), whereby population levels below a certain size $S$ are
unharvested, while above that size the harvest strategy aims to return
the population to $S$. Whenever a model predicts that the population
will not persist below a certain threshold, the optimal solution is to
harvest the entire population immediately, resulting in an escapement
$S=0$, as seen in the true (correct form, exact parameters) model, the
Allen model (correct form, estimated parameters) and the GP. Only the
structurally correct model (Allen model) and the GP produce policies
close to the true optimum policy.

\begin{figure}[htbp]
\centering
\includegraphics{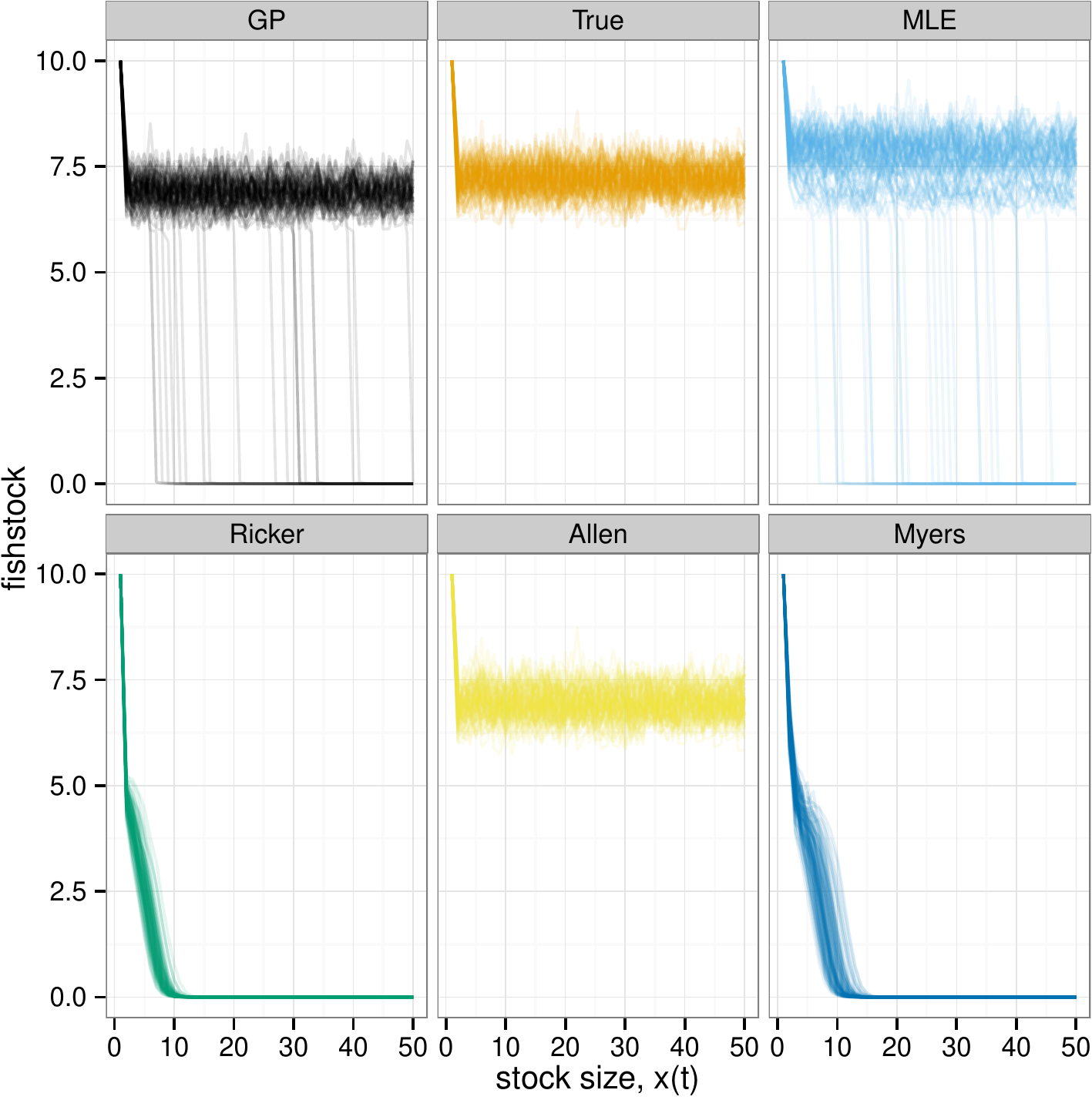}
\caption{In the management context, GPDP outperforms approaches based on
parametric models. We show 100 replicate simulations of the stock
dynamics (Eqn 1) under the policies derived from each of the estimated
models, as well as the policy based on the exact underlying model.}
\end{figure}

In Figure 4, we show the consequences of managing 100 replicate
realizations of the simulated fishery under policies derived from each
model. The structurally correct model under-harvests, leaving the stock
to vary around its unfished optimum. The structurally incorrect Ricker
model over-harvests the population past the tipping point consistently,
resulting in the immediate crash of the stock and thus leads to minimal
long-term catch.

The results across replicate stochastic simulations are most easily
compared by using the relative differences in net present value realized
by each of the model (Figure 5). Although not perfect, the GPDP
consistently realizes a value close to the optimal solution, and avoids
ever driving the system across the tipping point, which results in the
near-zero value cases in the parametric models.

\begin{figure}[htbp]
\centering
\includegraphics{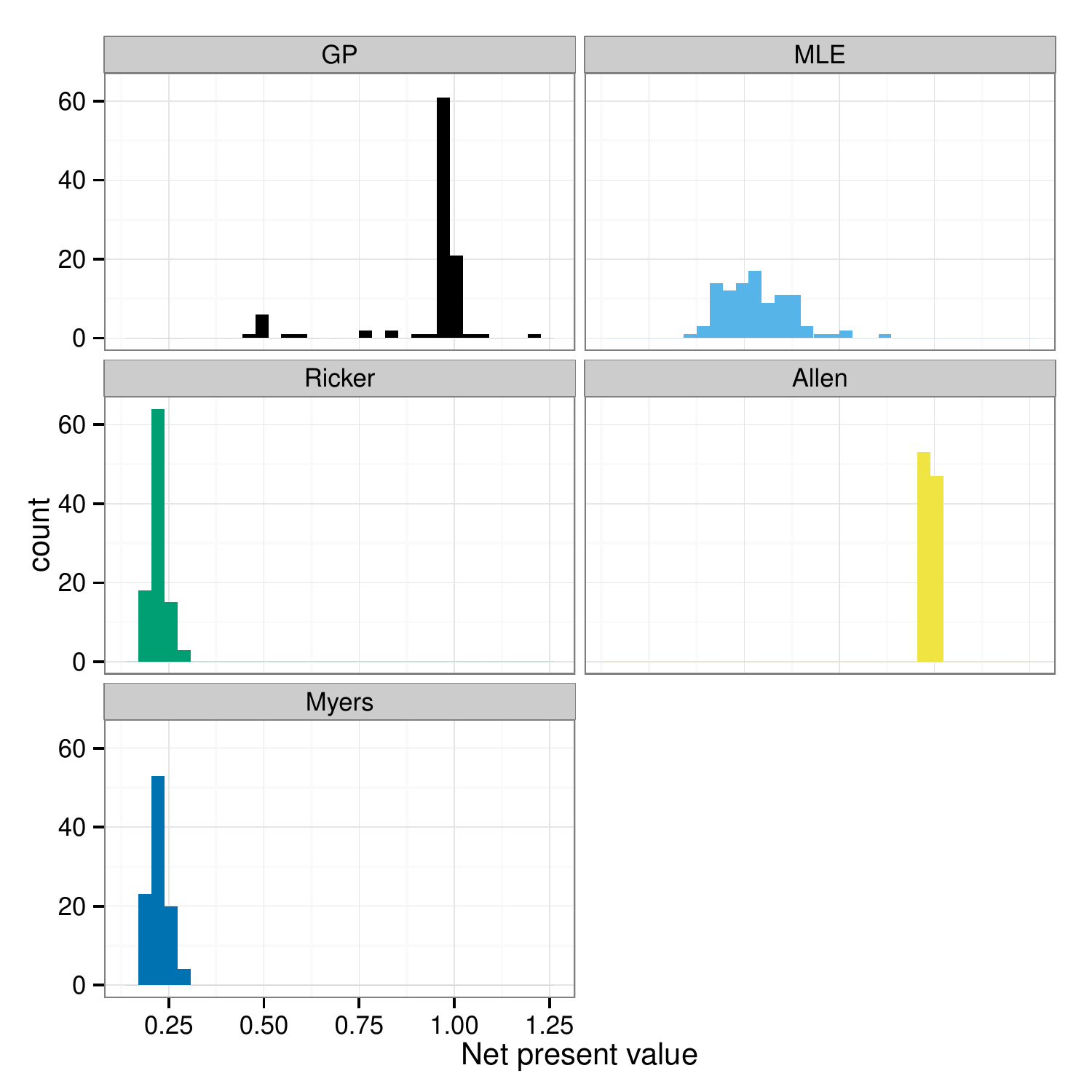}
\caption{Histograms of the realized net present value of the fishery
over a range of simulated data and resulting parameter estimates. For
each data set, the three models are estimated as described above.}
\end{figure}

\subsection{Sensitivity Analysis}\label{sensitivity-analysis}

These results are not sensitive to the modeling details of the
simulation. The GPDP estimate remains very close to the optimal solution
(obtained by knowing the true model) across changes to the training
simulation, scale of stochasticity, parameters or structure of the
underlying model. In the Supplement, we consider both a Latin hypercube
approach and a more focused investigation of the effects of the relative
distance to the Allee threshold and the variance of process
stochasticity.

The GPDP is only weakly influenced by increasing stochasticity or
increasing Allee effects over much of the range (Figure S2). Larger
$\sigma$ or higher Allee levels make even the optimal solution without
any model or parameter uncertainty unable to harvest the population
effectively (e.g.~the stochasticity is large enough to violate the
self-sustaining criterion of Reed (1979)).

\section{Discussion}\label{discussion}

Simple, mechanistically motivated models offer the potential to increase
our basic understanding of ecological processes (Geritz and Kisdi 2012,
Cuddington et al. 2013). But such models can be both inaccurate and
misleading when used in a decision making framework. In this paper we
tackled two aspects of uncertainty that are common to many ecological
decision-making problems and fundamentally challenging to existing
approaches that largely rely on parametric models:

\begin{enumerate}
\def\labelenumi{\arabic{enumi}.}
\itemsep1pt\parskip0pt\parsep0pt
\item
  We do not know the correct models for ecological systems.
\item
  We have limited data from which to estimate the model.
\end{enumerate}

We have illustrated how the use of non-parametric methods provides more
reliable solutions in the sequential decision-making problem.

\subsubsection{Traditional model-choice approaches can be positively
misleading}\label{traditional-model-choice-approaches-can-be-positively-misleading}

Our results illustrate that model-choice approaches can be absolutely
misleading -- by providing support to models that cannot capture tipping
point dynamics because they have fewer parameters and the data are far
from the tipping point. That is, when the data come from around the
stable steady state, all the parametric models are approximately linear
and approximately identical. Thus, it is intuitive that all model
selection methods choose the simplest model. In a complex world, the
result is suboptimal. But in a world that might contain tipping points,
the result could be disastrous.

Many managers both in fisheries and beyond face a similar situation:
they have not observed the population dynamics at all possible
densities. A lack of comprehensive data at all population sizes,
combined with the inability to formulate accurate population models for
low population sizes in the absence of data, makes this situation the
rule more than the exception. Relying on parametric models and model
choice processes that favor simplicity ignores this basic reality. For a
long time, Carl Walters (e.g. Walters and Hilborn 1978) has argued that
if we began by fishing any newly exploited population down to very low
levels and then let it recover, we would be much better at estimating
population dynamics and thus predicting the optimal harvest levels.
While certainly true, this presents a rather risky policy in the face of
potential tipping points. The GPDP offers a risk-adverse alternative.

\subsubsection{GPDP population dynamics capture larger uncertainty in
regions where the data are
poor}\label{gpdp-population-dynamics-capture-larger-uncertainty-in-regions-where-the-data-are-poor}

Parametric models perform most poorly when we seek a management strategy
outside the range of the observed data. The GPDP, in contrast, leads to
a predictive model that expresses a great deal of uncertainty about the
probable dynamics \emph{outside} the range of the observed data, while
retaining very good predictive accuracy \emph{inside} the range. The
management policy based on by the GPDP balances uncertainty outside the
range of the observed data against the immediate value of the harvest,
and acts to stabilize the population dynamics in a region of state space
in which the predictions are reliably reflected by the data.

Such problems are ubiquitous across ecological decision-making and
conservation where the greatest concerns involve decisions that lead to
population sizes that have never been observed and for which we do not
know the response -- whether this is the collapse of a fishery, the
spread of an invasive, or the loss of habitat.

\subsubsection{The role of the prior}\label{the-role-of-the-prior}

Outside of the observed range of the data, the GP reverts to the prior,
and consequently the choice of the prior can also play a significant
role in determining the optimal policy. In the examples shown here we
have selected a prior that is both relatively uninformative (due to the
broad priors placed on its parameters $\ell$ and $\sigma$ and simple
(the choice of our covariance function, Eqns 12 and 13 ). In practice,
these should be chosen to confer particular biological properties. In
principle, this may allow a manager to improve the performance of the
GPDP by adding detail as is justified. For instance, it would be
possible to use a linear or a Ricker-shaped mean in the prior without
making the much stronger assumption that the Ricker is the structurally
correct model (Sugeno and Munch 2013a). One fruitful avenue of future
research is identifying criteria to ensure the prior and the reward
function are chosen appropriately for the problem at hand.

\section{Acknowledgments}\label{acknowledgments}

This work was partially supported by NOAA-IAM grant to SM and Alec
McCall and administered through the Center for Stock Assessment
Research, a partnership between the University of California Santa Cruz
and the Fisheries Ecology Division, Southwest Fisheries Science Center,
Santa Cruz, CA and by NSF grant EF-0924195 to MM and NSF grant
DBI-1306697 to CB.

Allen, D., and K. Tanner. 2005. Infusing active learning into the
large-enrollment biology class: seven strategies, from the simple to
complex. Cell biology education 4:262--8.

Athanassoglou, S., and A. Xepapadeas. 2012. Pollution control with
uncertain stock dynamics: When, and how, to be precautious. Journal of
Environmental Economics and Management 63:304--320.

Boettiger, C., M. Mangel, and S. Munch. 2014. nonparametric-bayes
v0.1.0. ZENODO. \url{http://dx.doi.org/10.5281/zenodo.12669}.

Brozović, N., and W. Schlenker. 2011. Optimal management of an ecosystem
with an unknown threshold. Ecological Economics:1--14.

Burnham, K. P., and D. R. Anderson. 2002. Model Selection and
Multi-Model Inference. Page 496. Springer.

Clark, C. W. 1976. Mathematical Bioeconomics. WileyNew York.

Clark, C. W. 2009. Mathematical Bioeconomics. WileyNew York.

Clark, C. W., and G. P. Kirkwood. 1986. On uncertain renewable resource
stocks: Optimal harvest policies and the value of stock surveys. Journal
of Environmental Economics and Management 13:235--244.

Clark, C. W., and M. Mangel. 2000. Dynamic state variable models in
ecology. Oxford University PressOxford.

Courchamp, F., L. Berec, and J. Gascoigne. 2008. Allee Effects in
Ecology and Conservation. Page 256. Oxford University Press, USA.

Cressie, N., C. a Calder, J. S. Clark, J. M. {Ver Hoef}, and C. K.
Wikle. 2009. Accounting for uncertainty in ecological analysis: the
strengths and limitations of hierarchical statistical modeling.
Ecological Applications 19:553--70.

Cuddington, K. M., M. Fortin, and L. Gerber. 2013. Process-based models
are required to manage ecological systems in a changing world. Ecosphere
4:1--12.

Fischer, J., G. D. Peterson, T. a Gardner, L. J. Gordon, I. Fazey, T.
Elmqvist, A. Felton, C. Folke, and S. Dovers. 2009. Integrating
resilience thinking and optimisation for conservation. Trends in ecology
\& evolution 24:549--54.

Gelman, A., J. B. Carlin, H. S. Stern, and D. B. Rubin. 2003. Bayesian
Data Analysis. 2nd editions. Chapman; Hall/CRC.

Geritz, S. A. H., and E. Kisdi. 2012. Mathematical ecology: why
mechanistic models? Journal of mathematical biology 65:1411--5.

Gordon, H. 1954. The economic theory of a common-property resource: the
fishery. The Journal of Political Economy 62:124--142.

Hilborn, R., and M. Mangel. 1997. The Ecological Detective: Confronting
Models with data. Page 330. Princeton University Press.

Hughes, T. P., C. Linares, V. Dakos, I. a van de Leemput, and E. H. van
Nes. 2013. Living dangerously on borrowed time during slow, unrecognized
regime shifts. Trends in ecology \& evolution 28:149--55.

Kocijan, J., A. Girard, B. Banko, and R. Murray-Smith. 2005. Dynamic
systems identification with Gaussian processes. Mathematical and
Computer Modelling of Dynamical Systems 11:411--424.

Ludwig, D., and C. J. Walters. 1982. Optimal harvesting with imprecise
parameter estimates. Ecological Modelling 14:273--292.

Mangel, M. 2014. Stochastic Dynamic Programming Illuminates the Link
Between Environment. Bulletin of Mathematical Biology in press.

Mangel, M., and C. W. Clark. 1988. Dynamic Modeling in Behavioral
Ecology. (J. Krebs and T. Clutton-Brock, Eds.). Princeton University
PressPrinceton.

Mangel, M., 0. Fiksen, and J. Giske. 2001. Theoretical and statistical
models in natural resource management and research. Pages 57--71
\emph{in} T. M. Shenk and A. B. Franklin, editors. Modeling in natural
resource management, development, interpretation and application. Island
PressWashington DC.

Marescot, L., G. Chapron, I. Chadès, P. L. Fackler, C. Duchamp, E.
Marboutin, and O. Gimenez. 2013. Complex decisions made simple: a primer
on stochastic dynamic programming. Methods in Ecology and
Evolution:n/a--n/a.

May, R. M., J. R. Beddington, C. W. Clark, S. J. Holt, and R. M. Laws.
1979. Management of multispecies fisheries. Science (New York, N.Y.)
205:267--77.

McAllister, M. 1998. Bayesian stock assessment: a review and example
application using the logistic model. ICES Journal of Marine Science
55:1031--1060.

Munch, S. B., A. Kottas, and M. Mangel. 2005a. Bayesian nonparametric
analysis of stock-recruitment relationships. Canadian Journal of
Fisheries and Aquatic Sciences 62:1808--1821.

Munch, S. B., M. L. Snover, G. M. Watters, and M. Mangel. 2005b. A
unified treatment of top-down and bottom-up control of reproduction in
populations. Ecology Letters 8:691--695.

Myers, R. A., N. J. Barrowman, J. A. Hutchings, and A. a Rosenberg.
1995. Population dynamics of exploited fish stocks at low population
levels. Science (New York, N.Y.) 269:1106--8.

Polasky, S., S. R. Carpenter, C. Folke, and B. Keeler. 2011.
Decision-making under great uncertainty: environmental management in an
era of global change. Trends in ecology \& evolution:1--7.

R Core Team. 2013. R: A Language and Environment for Statistical
Computing. R Foundation for Statistical ComputingVienna, Austria.

Rasmussen, C. E., and C. K. I. Williams. 2006. Gaussian Processes for
Machine Learning. (Thomas Dietterich, Ed.). MIT Press,Boston.

Reed, W. J. 1979. Optimal escapement levels in stochastic and
deterministic harvesting models. Journal of Environmental Economics and
Management 6:350--363.

Roughgarden, J. E., and F. Smith. 1996. Why fisheries collapse and what
to do about it. Proceedings of the National Academy of Sciences of the
United States of America 93:5078.

Schapaugh, A. W., and A. J. Tyre. 2013. Accounting for parametric
uncertainty in Markov decision processes. Ecological Modelling
254:15--21.

Scheffer, M., J. Bascompte, W. A. Brock, V. Brovkin, S. R. Carpenter, V.
Dakos, H. Held, E. H. van Nes, M. Rietkerk, and G. Sugihara. 2009.
Early-warning signals for critical transitions. Nature 461:53--9.

Scheffer, M., S. R. Carpenter, J. A. Foley, C. Folke, and B. Walker.
2001. Catastrophic shifts in ecosystems. Nature 413:591--6.

Sethi, G., C. Costello, A. Fisher, M. Hanemann, and L. Karp. 2005.
Fishery management under multiple uncertainty. Journal of Environmental
Economics and Management 50:300--318.

Sigourney, D. B., S. B. Munch, and B. H. Letcher. 2012. Combining a
Bayesian nonparametric method with a hierarchical framework to estimate
individual and temporal variation in growth. Ecological Modelling
247:125--134.

Su, Y.-S., and Masanao Yajima. 2013. R2jags: A Package for Running jags
from R.

Sugeno, M., and S. B. Munch. 2013a. A semiparametric Bayesian method for
detecting Allee effects. Ecology 94:1196--1204.

Sugeno, M., and S. B. Munch. 2013b. A semiparametric Bayesian approach
to estimating maximum reproductive rates at low population sizes.
Ecological applications : a publication of the Ecological Society of
America 23:699--709.

Thorson, J. T., K. Ono, and S. B. Munch. 2014. A Bayesian approach to
identifying and compensating for model misspecification in population
models. Ecology 95:329--41.

Walters, C. J., and R. Hilborn. 1978. Ecological Optimization and
Adaptive Management. Annual Review of Ecology and Systematics
9:157--188.

Weitzman, M. L. 2013. A Precautionary Tale of Uncertain Tail Fattening.
Environmental and Resource Economics 55:159--173.

Williams, B. K. 2001. Uncertainty , learning , and the optimal
management of wildlife. Environmental and Ecological Statistics
8:269--288.

Xie, Y. 2013. Dynamic Documents with R and knitr. Chapman; Hall/CRCBoca
Raton, Florida.

\appendix
\renewcommand*{\thefigure}{S\arabic{figure}}
\renewcommand*{\thetable}{S\arabic{table}} \setcounter{figure}{0}
\setcounter{table}{0}

\section{Code}\label{code}

All code used in producing this analysis has been embedded into the
manuscript sourcefile using the Dynamic Documentation tool,
\texttt{knitr} for the R language (Xie 2013), available at
\href{https://github.com/cboettig/nonparametric-bayes/}{github.com/cboettig/nonparametric-bayes/}

To help make the mathematical and computational approaches presented
here more accessible, we provide a free and open source (MIT License) R
package that implements the GPDP process as it is presented here. Users
should note that at this time, the R package has been developed and
tested explicitly for this analysis and is not yet intended as a general
purpose tool. The manuscript source-code described above illustrates how
these functions are used in this analysis. This package can be installed
following the directions above.

\subsection{Dependencies \&
Reproducibility}\label{dependencies-reproducibility}

The code provided should run on any common platform (Windows, Mac or
Linux) that has R and the necessary R packages installed (including
support for the jags Gibbs sampler). The DESCRIPTION file of our R
package, \texttt{nonparametricbayes}, lists all the software required to
use these methods. Additional software requirements for the other
comparisons shown here, such as the Gibbs sampling for the parametric
models, are listed under the Suggested packages list.

Nonetheless, installing the dependencies needed is not a trivial task,
and may become more difficult over time as software continues to evolve.
To facilitate reuse, we also provide a Dockerfile and Docker image that
can be used to replicate and explore the analyses here by providing a
copy of the computational environment we have used, with all software
installed. Docker sofware (see \href{http://www.docker.com}{docker.com})
runs on most platforms as well as cloud servers. Use the command:

\begin{verbatim}
docker run -dP cboettig/nonparametric-bayes
\end{verbatim}

to launch an RStudio instance with the necessary software already
installed. See the Rocker-org project,
\href{https://github.com/rocker-org}{github.com/rocker-org} for more
detailed documentation on using Docker with R.

\section{Data}\label{data}

\subsection{Dryad Data Archive}\label{dryad-data-archive}

While the data can be regenerated using the code provided, for
convenience CSV files of the data shown in each graph are made available
on Dryad, along with the source \texttt{.Rmd} files for the manuscript
and supplement that document them.

\subsection{Training data description}\label{training-data-description}

Each of our models $f(S_t)$ must be estimated from training data, which
we simulate from the Allen model with parameters $r = $ 2, $K =$ 8,
$C =$ 5, and $\sigma_g =$ 0.05 for $T=$ 40 timesteps, starting at
initial condition $X_0 = $ 5.5. The training data can be seen in Figure
1 and found in the table \texttt{figure1.csv}.

\subsection{Training data for sensitivity
analyses}\label{training-data-for-sensitivity-analyses}

A further 96 unique randomly generated training data sets are generated
for the sensitivity analysis, as described in the main text. The code
provided replicates the generation of these sets.

\section{Model performance outside the predicted range (Fig
S1)}\label{model-performance-outside-the-predicted-range-fig-s1}

Figure S1 illustrates the performance of the GP and parametric models
outside the observed training data. The mean trajectory under the
underlying model is shown by the black dots, while the corresponding
prediction made by the model shown by the box and whiskers plots.
Predictions are based on the true expected value in the previous time
step. Predicted distributions that lie entirely above the expected
dynamics indicate the expectation of stock sizes higher than what is
actually expected. The models differ both in their expectations and
their uncertainty (colored bands show two standard deviations away).
Note that the GP is particularly uncertain about the dynamics relative
to structurally incorrect models like the Ricker.

\begin{figure}[htbp]
\centering
\includegraphics{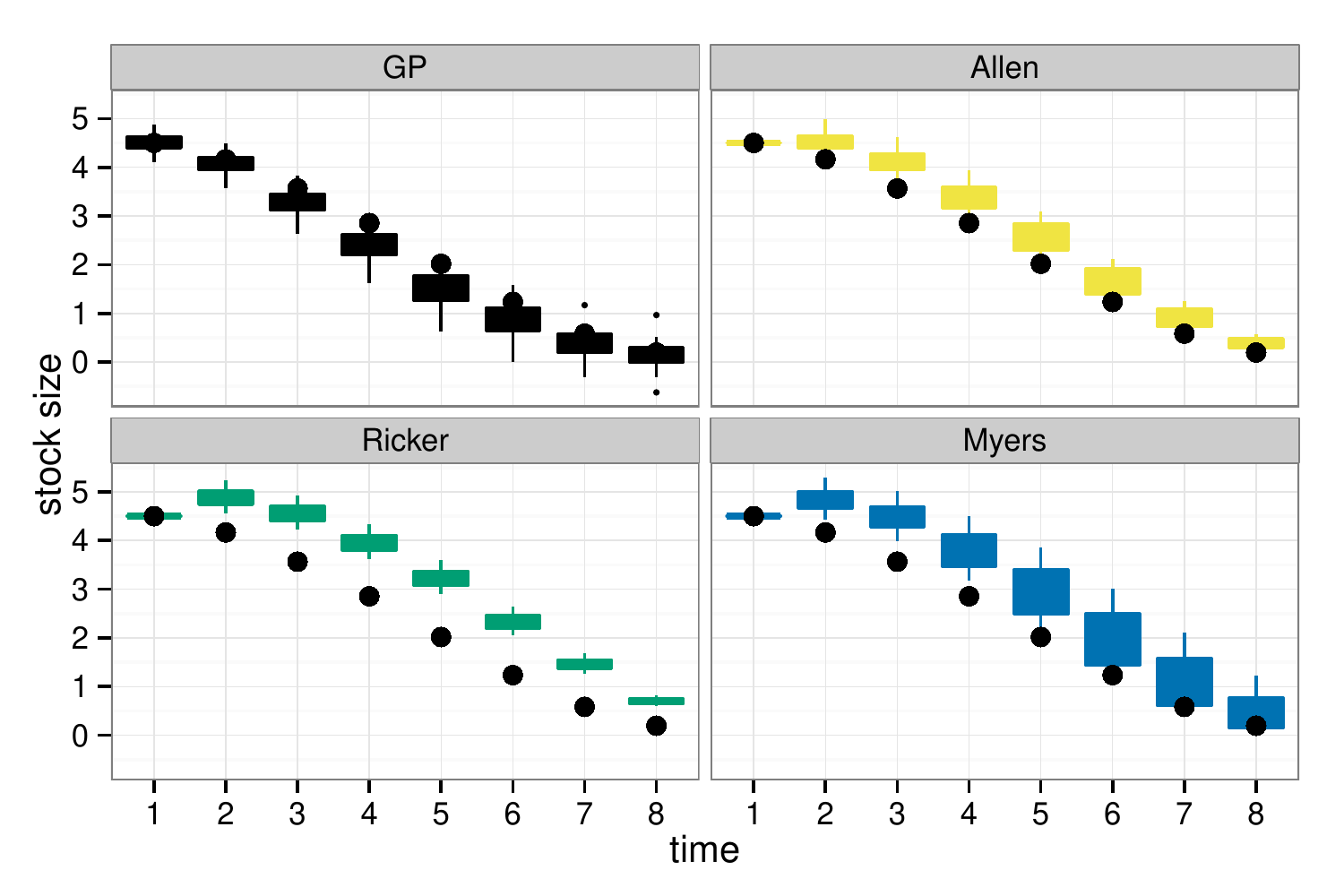}
\caption{Outside the range of the training data (Figure 1), the true
dynamics (black dots) fall outside the uncertainty (two standard
deviations, colored bands) of the structurally incorrect parametric
models (Ricker, Myers), but inside the uncertainty predicted by the GP.
Points show the stock size simulated by the true model. Overlay shows
the range of states predicted by each model, based on the state observed
in the previous time step. The Ricker model always (wrongly) predicts
positive population growth, while the actual population shrinks in each
step as the initial condition falls below the Allee threshold of the
underlying model (Allen). Note that because it does not assume a
parametric form but instead relies more directly on the data, the GP is
both more pessimistic and more uncertain about the future state than the
parametric models.}
\end{figure}

\newpage

\section{Further sensitivity analysis (Fig S2 -
3)}\label{further-sensitivity-analysis-fig-s2---3}

We perform 2 sensitivity analyses. The first focuses on illustrating the
robustness of the approach to the two parameters that most influence
stochastic transitions across the tipping point: the position of the
Allee threshold and the scale of the noise (Fig S2).

Changing the intensity of the stochasticity or the distance between
stable and unstable steady states does not impact the performance of the
GP relative to the optimal solution obtained from the true model and
true parameters. The parametric models are more sensitive to this
difference. Large values of $\sigma$ relative to the distance between
the stable and unstable point increases the chance of a stochastic
transition below the tipping point. More precisely, if we let $L$ be the
distance between the stable and unstable steady states, then the
probability that fluctuations drive the population across the unstable
steady state scales as

$\exp\left(\frac{L^2}{\sigma^2}\right)$

(see Gardiner (2009) or Mangel (2006) for the derivation).

Thus, the impact of using a model that underestimates the risk of
harvesting beyond the critical point is considerable, since this such a
situation occurs more often. Conversely, with large enough distance
between the optimal escapement and unstable steady state relative to
$\sigma$, the chance of a transition becomes vanishingly small and all
models can be estimated near-optimally. Models that underestimate the
cost incurred by population sizes fluctuating significantly below the
optimal escapement level will not perform poorly as long as those
fluctuations are sufficiently small. Fig S2 shows the net present value
of managing under teh GPDP remains close to the optimal value (ratio of
1), despite varying across either noise level or the the position of the
allee threshold.

\begin{figure}[htbp]
\centering
\includegraphics{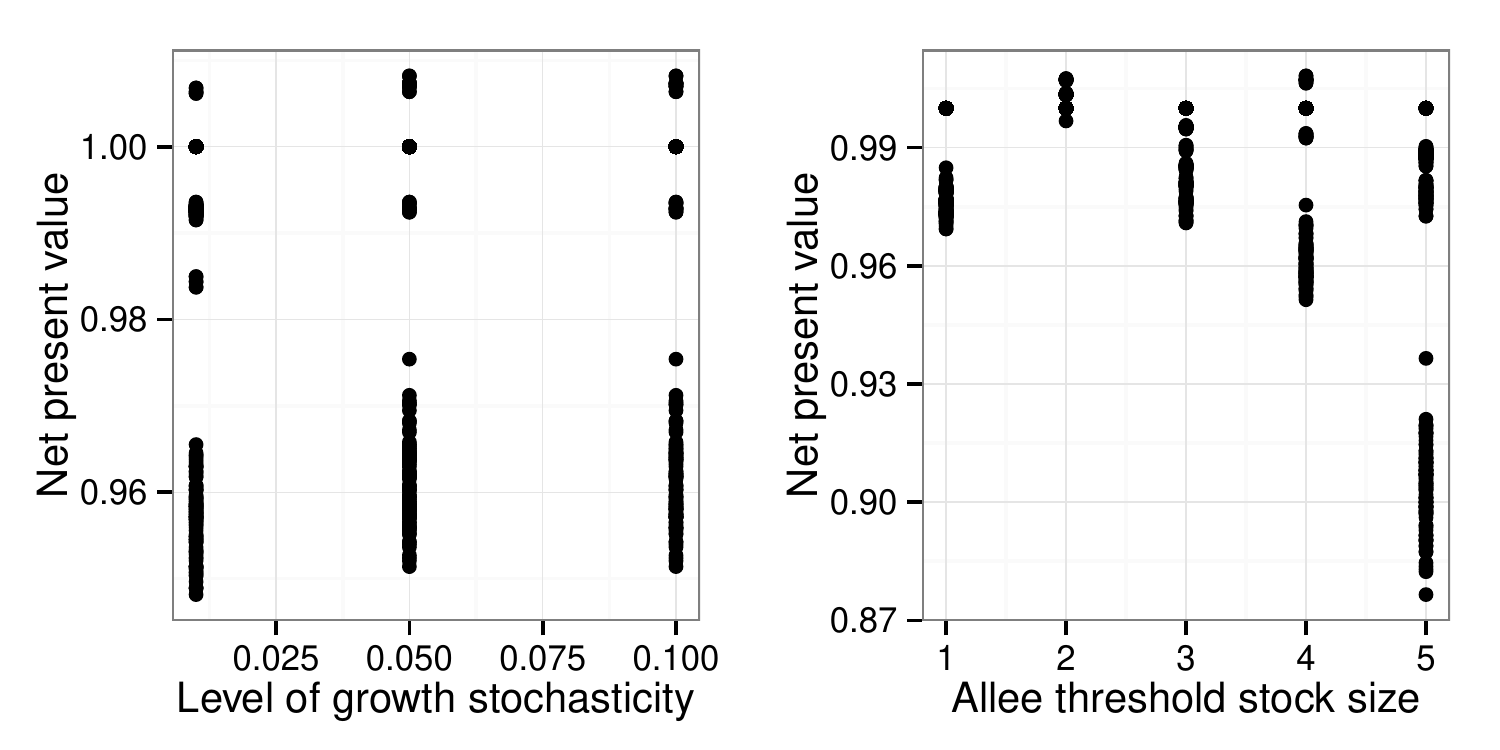}
\caption{The effect of increasing noise or decreasing Allee threshold
levels on the net present value of the fishery when managed under the
GPDP, relative to managing under the true model (with known parameters).
Other than the focal parameter (stochasticity, Allee threshold), other
parameters are held fixed as above to illustrate this effect.}
\end{figure}

The Latin hypercube approach systematically varies all combinations of
parameters, providing a more general test than varying only one
parameter at a time. We loop across eight replicates of three different
randomly generated parameter sets for each of two different generating
models (Allen and Myers) over two different noise levels (0.01 and
0.05), for a total of 8 x 3 x 2 x 2 = 96 scenarios. The Gaussian Process
performs nearly optimally in each case, relative to the optimal solution
with no parameter or model uncertainty (Figure S10, appendix).

\begin{figure}[htbp]
\centering
\includegraphics{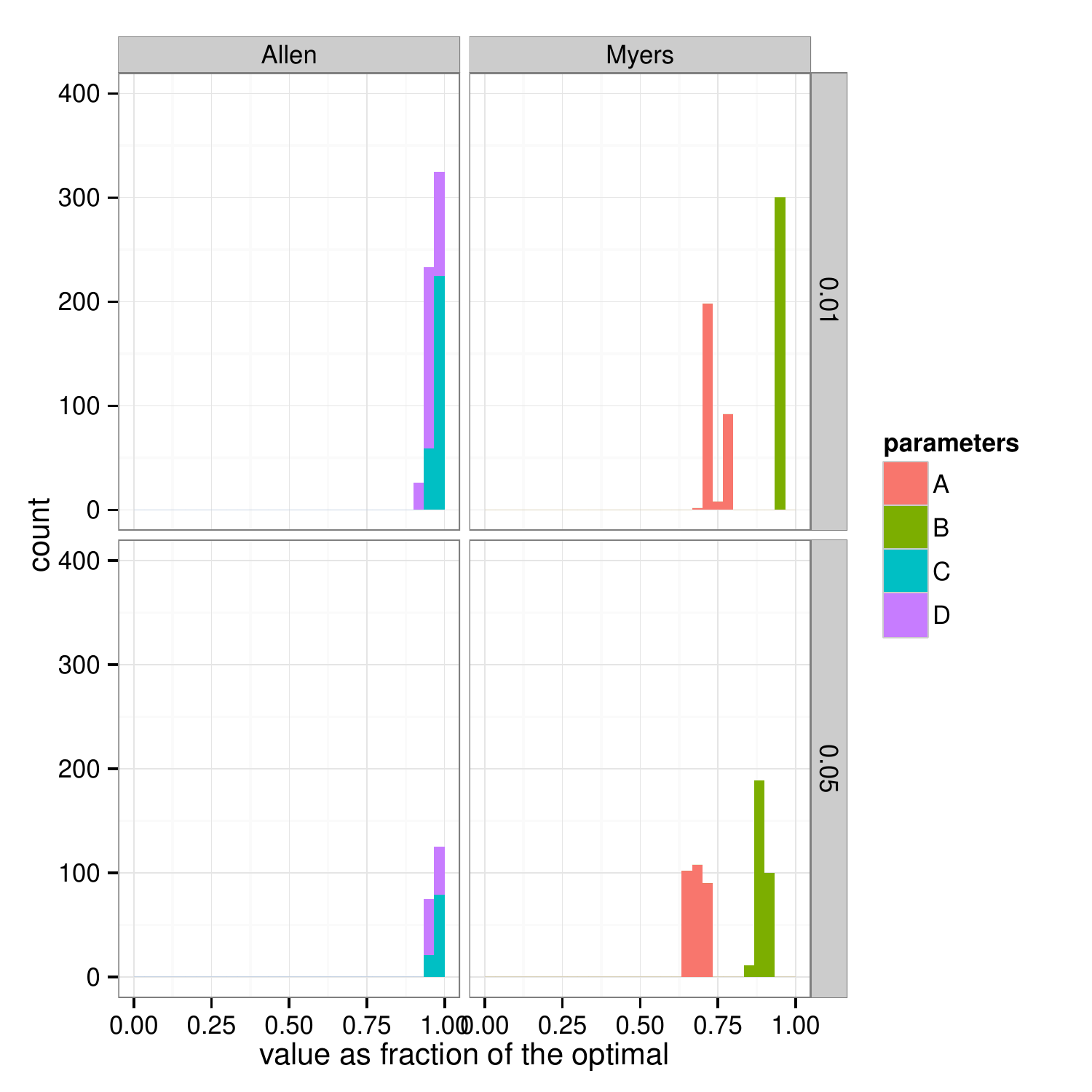}
\caption{Sensitivity Analysis. Histograms shows the ratio of the
realized net present value derived when managing under the GPDP over the
optimal value given the true model and true parameters. Values of 1
indicate optimal performance. Columns indicate different models, rows
different noise levels, and colors indicate the parameter set used.
Grouped over stochastic replicates applying the contol policy and
stochastic replicates of training data generated from the model
indicated, see raw data for details. Randomly chosen parameter values
for the models shown in tables below.}
\end{figure}

\begin{longtable}[c]{@{}cccc@{}}
\toprule\addlinespace
\begin{minipage}[b]{0.15\columnwidth}\centering
~
\end{minipage} & \begin{minipage}[b]{0.07\columnwidth}\centering
r
\end{minipage} & \begin{minipage}[b]{0.07\columnwidth}\centering
K
\end{minipage} & \begin{minipage}[b]{0.09\columnwidth}\centering
theta
\end{minipage}
\\\addlinespace
\midrule\endhead
\begin{minipage}[t]{0.15\columnwidth}\centering
\textbf{set.A}
\end{minipage} & \begin{minipage}[t]{0.07\columnwidth}\centering
1.103
\end{minipage} & \begin{minipage}[t]{0.07\columnwidth}\centering
7.949
\end{minipage} & \begin{minipage}[t]{0.09\columnwidth}\centering
2.288
\end{minipage}
\\\addlinespace
\begin{minipage}[t]{0.15\columnwidth}\centering
\textbf{set.B}
\end{minipage} & \begin{minipage}[t]{0.07\columnwidth}\centering
1.485
\end{minipage} & \begin{minipage}[t]{0.07\columnwidth}\centering
9.775
\end{minipage} & \begin{minipage}[t]{0.09\columnwidth}\centering
3.524
\end{minipage}
\\\addlinespace
\bottomrule
\addlinespace
\caption{Randomly chosen parameter sets for the Allen models in Figure
S3.}
\end{longtable}

\begin{longtable}[c]{@{}cccc@{}}
\toprule\addlinespace
\begin{minipage}[b]{0.15\columnwidth}\centering
~
\end{minipage} & \begin{minipage}[b]{0.07\columnwidth}\centering
r
\end{minipage} & \begin{minipage}[b]{0.07\columnwidth}\centering
K
\end{minipage} & \begin{minipage}[b]{0.07\columnwidth}\centering
C
\end{minipage}
\\\addlinespace
\midrule\endhead
\begin{minipage}[t]{0.15\columnwidth}\centering
\textbf{set.C}
\end{minipage} & \begin{minipage}[t]{0.07\columnwidth}\centering
1.769
\end{minipage} & \begin{minipage}[t]{0.07\columnwidth}\centering
10.46
\end{minipage} & \begin{minipage}[t]{0.07\columnwidth}\centering
4.301
\end{minipage}
\\\addlinespace
\begin{minipage}[t]{0.15\columnwidth}\centering
\textbf{set.D}
\end{minipage} & \begin{minipage}[t]{0.07\columnwidth}\centering
2.075
\end{minipage} & \begin{minipage}[t]{0.07\columnwidth}\centering
10.95
\end{minipage} & \begin{minipage}[t]{0.07\columnwidth}\centering
4.915
\end{minipage}
\\\addlinespace
\bottomrule
\addlinespace
\caption{Randomly chosen parameter sets for the Myers models in Figure
S3.}
\end{longtable}

\newpage

\section{MCMC analyses}\label{mcmc-analyses}

This section provides figures and tables showing the traces from each of
the MCMC runs used to estimate the parameters of the models presented,
along with the resulting posterior distributions for each parameter.
Priors usually appear completely falt when shown against the posteriors,
but are summarized by tables the parameters of their corresponding
distributions for each case.

\subsection{GP MCMC (Fig S4-5)}\label{gp-mcmc-fig-s4-5}

\begin{figure}[htbp]
\centering
\includegraphics{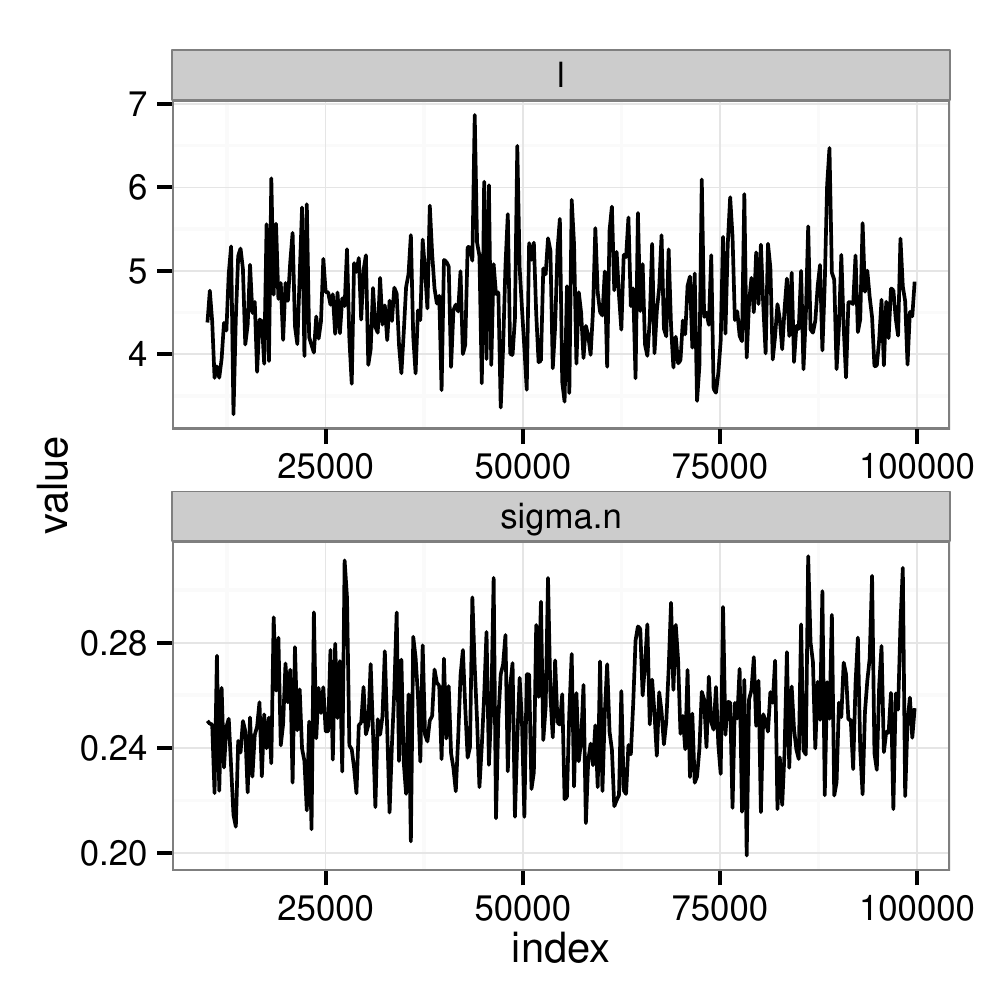}
\caption{Traces from the MCMC estimates of the GP model show reasonable
mixing (no trend) and sampling rejection rate (no piecewise jumps)}
\end{figure}

\begin{figure}[htbp]
\centering
\includegraphics{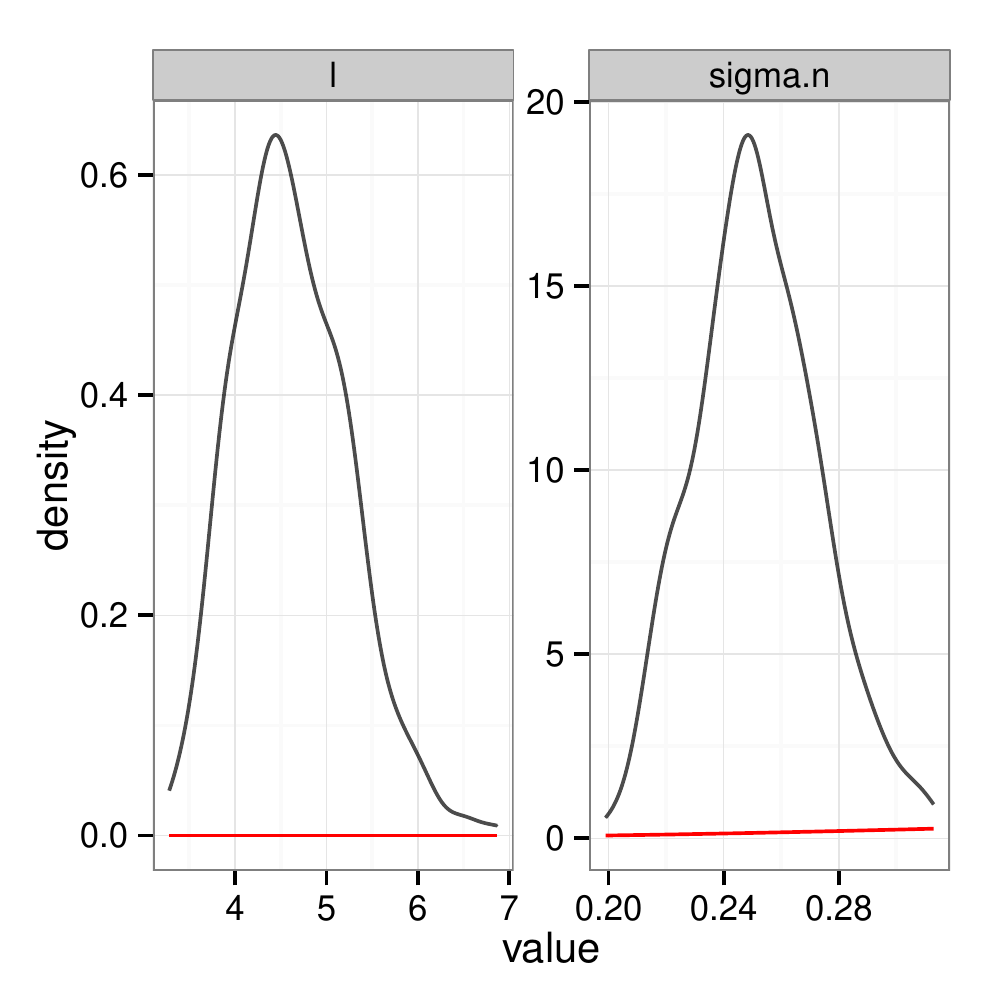}
\caption{Posterior distributions from the MCMC estimate of the GP model.
Prior curves shown in red; note the posterior distributions are
significantly more peaked than the priors, showing that the data has
been informative and is not driven by the priors.}
\end{figure}

\newpage
\newpage

\subsection{Ricker Model MCMC (Fig
S6-7)}\label{ricker-model-mcmc-fig-s6-7}

\begin{figure}[htbp]
\centering
\includegraphics{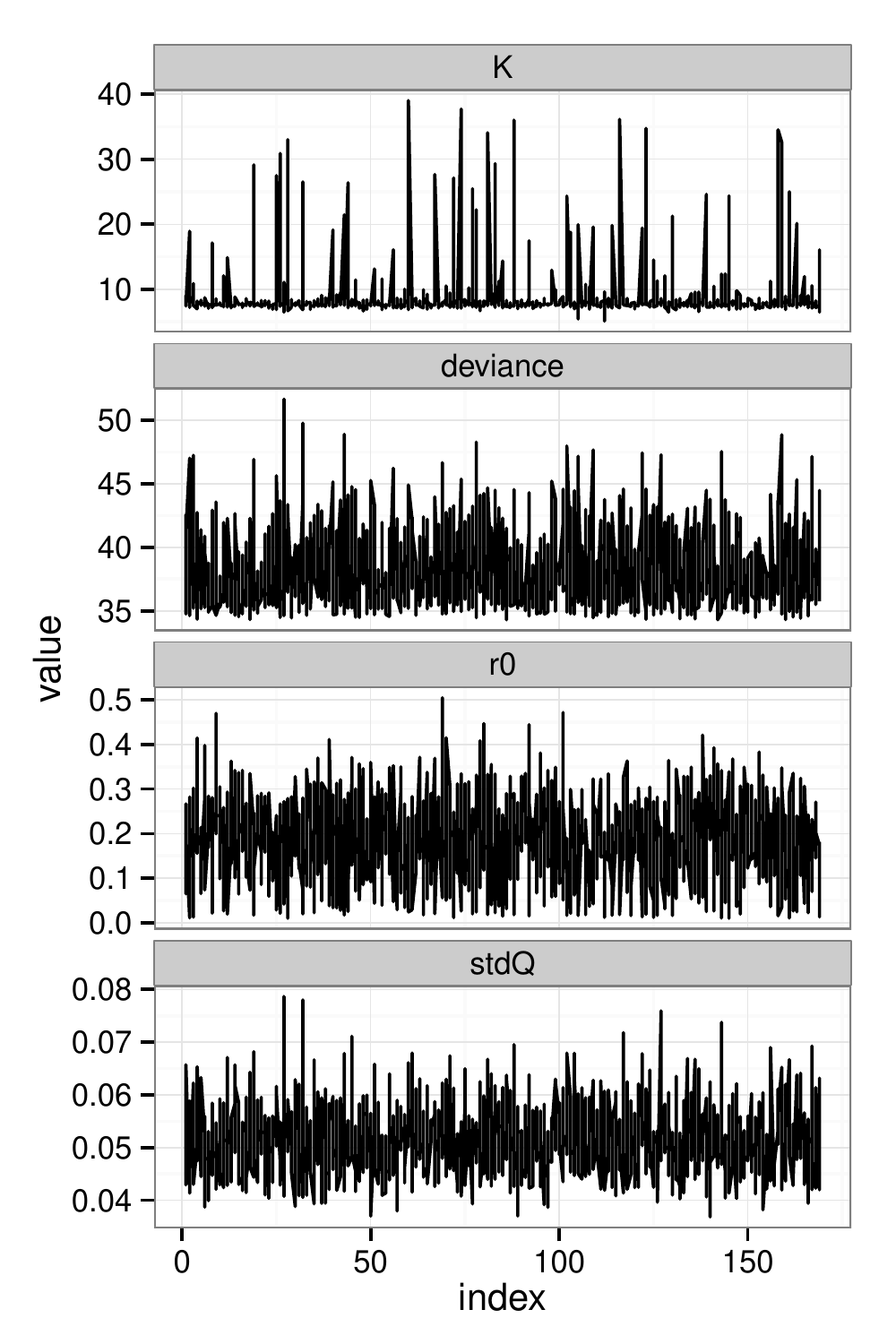}
\caption{Traces from the MCMC estimates of the Ricker model show
reasonable mixing (no trend) and sampling rejection rate (no piecewise
jumps). stdQ refers to the estimate of $\sigma$; deviance is -2 times
the log likelihood.}
\end{figure}

\begin{figure}[htbp]
\centering
\includegraphics{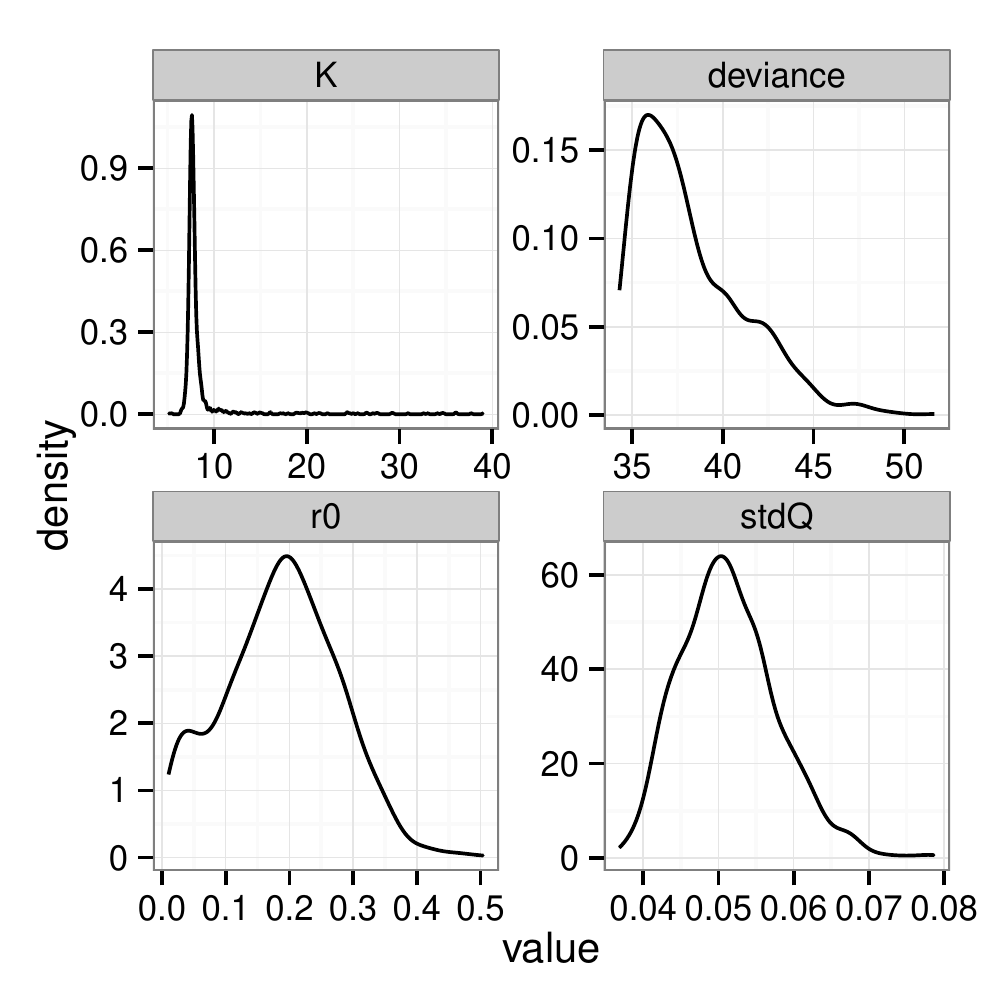}
\caption{Posteriors from the MCMC estimate of the Ricker model. Note
that the model estimates a carrying capacity $K$ very close to the true
equilibrium where most of the observations were made, but is less
certain about the growth rate.}
\end{figure}

\begin{longtable}[c]{@{}ccc@{}}
\toprule\addlinespace
\begin{minipage}[b]{0.15\columnwidth}\centering
parameter
\end{minipage} & \begin{minipage}[b]{0.18\columnwidth}\centering
lower.bound
\end{minipage} & \begin{minipage}[b]{0.18\columnwidth}\centering
upper.bound
\end{minipage}
\\\addlinespace
\midrule\endhead
\begin{minipage}[t]{0.15\columnwidth}\centering
$r$
\end{minipage} & \begin{minipage}[t]{0.18\columnwidth}\centering
0.01
\end{minipage} & \begin{minipage}[t]{0.18\columnwidth}\centering
20
\end{minipage}
\\\addlinespace
\begin{minipage}[t]{0.15\columnwidth}\centering
$K$
\end{minipage} & \begin{minipage}[t]{0.18\columnwidth}\centering
0.01
\end{minipage} & \begin{minipage}[t]{0.18\columnwidth}\centering
40
\end{minipage}
\\\addlinespace
\begin{minipage}[t]{0.15\columnwidth}\centering
$\sigma$
\end{minipage} & \begin{minipage}[t]{0.18\columnwidth}\centering
1e-06
\end{minipage} & \begin{minipage}[t]{0.18\columnwidth}\centering
100
\end{minipage}
\\\addlinespace
\bottomrule
\addlinespace
\caption{Parameterization range for the uniform priors in the Ricker
model}
\end{longtable}

\newpage
\newpage 

\subsection{Myers Model MCMC (Fig
S8-9)}\label{myers-model-mcmc-fig-s8-9}

\begin{figure}[htbp]
\centering
\includegraphics{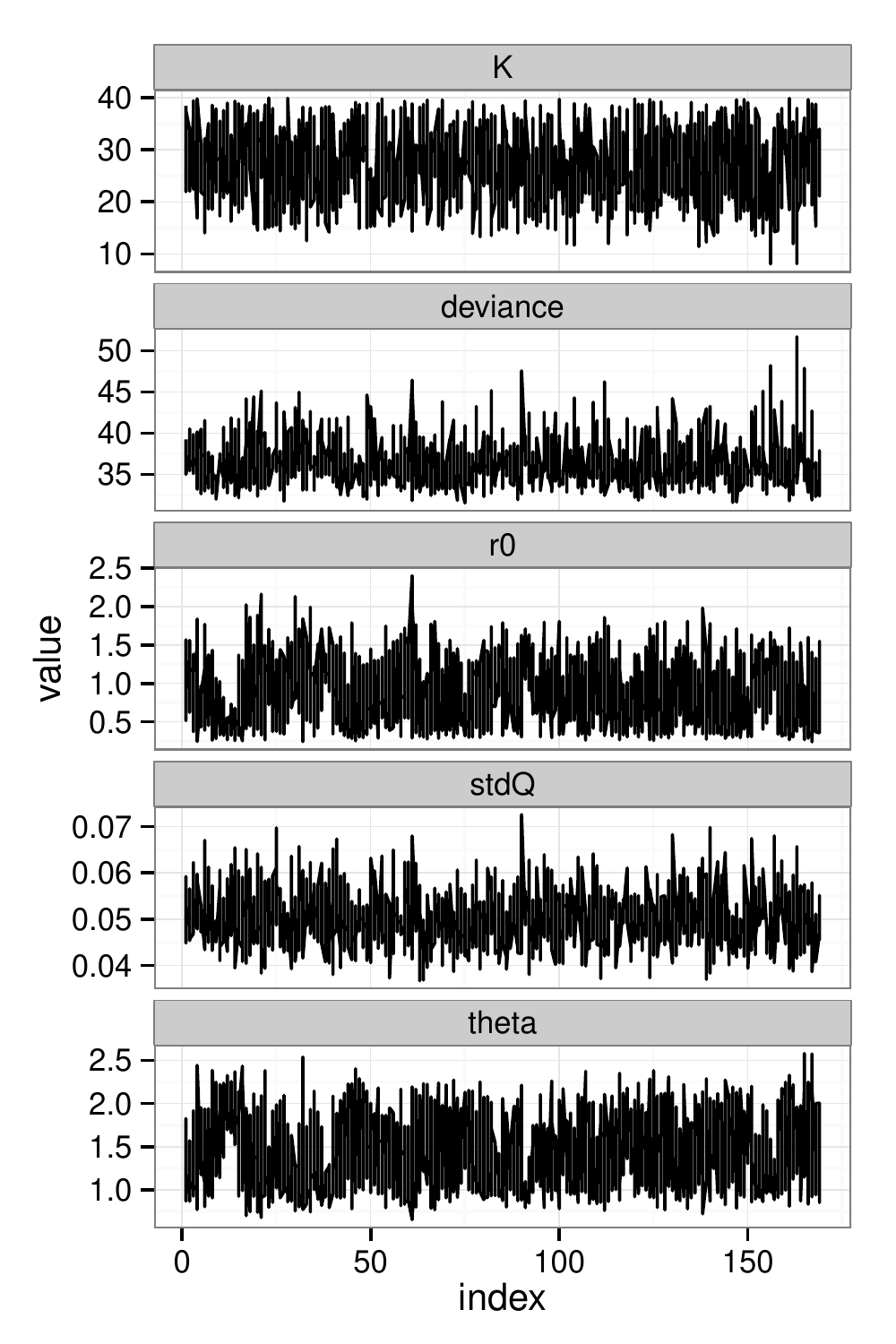}
\caption{Traces from the MCMC estimates of the Myers model show
reasonable mixing (no trend) and sampling rejection rate (no piecewise
jumps). stdQ refers to the estimate of $\sigma$; deviance is -2 times
the log likelihood.}
\end{figure}

\begin{figure}[htbp]
\centering
\includegraphics{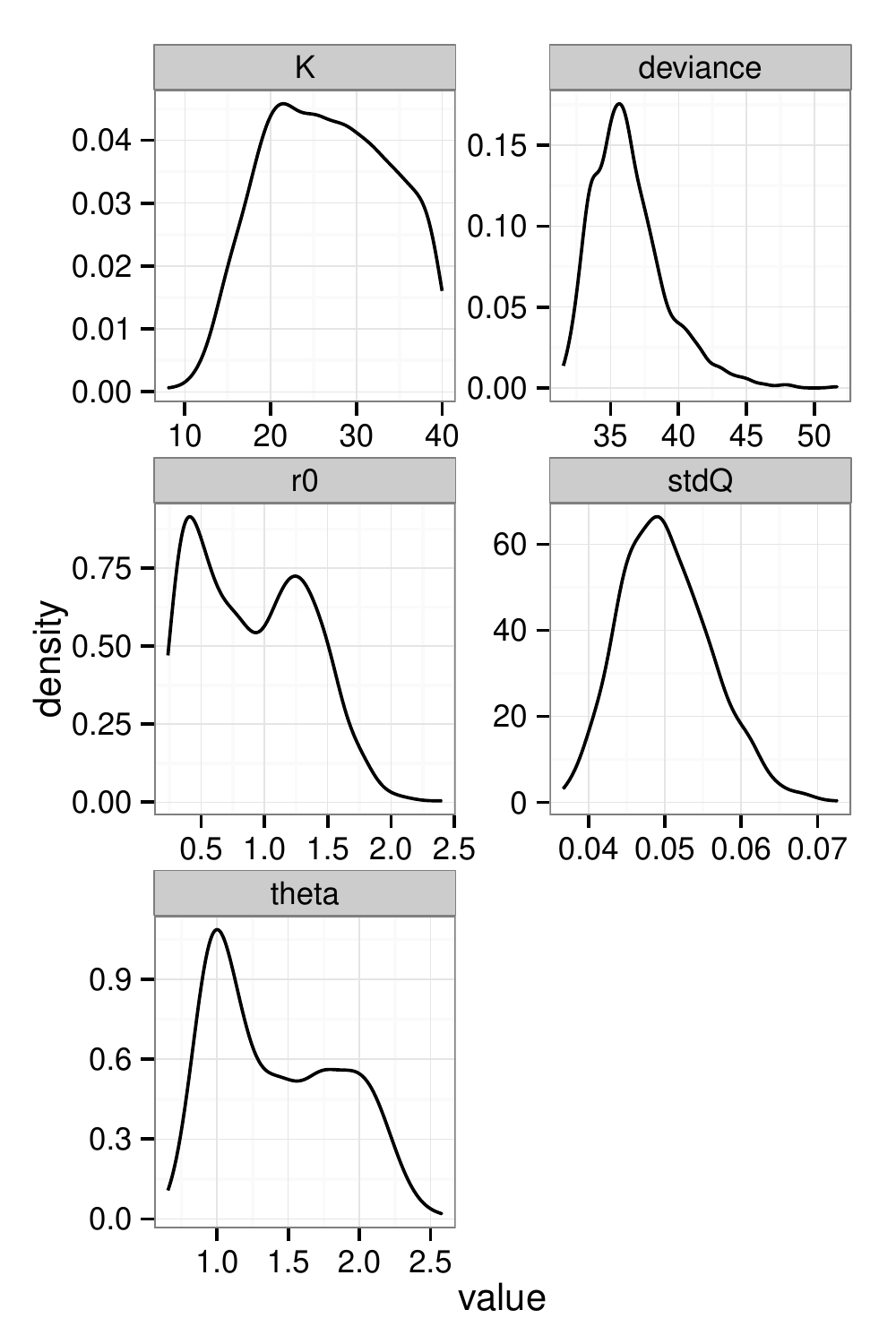}
\caption{Posterior distributions from the MCMC estimates of the Myers
model. Note that with more free parameters, the posteriors reflect
greater uncertainty. In particular, the parameter $\theta$ includes
values both above 2, resulting in a tipping point, and below 2, where no
tipping point exists in the model. Though the dynamic program will
integrate over the full distribution, including those values
corresponding to tipping points, the weight of the model lies in the
region without tipping points.}
\end{figure}

\begin{longtable}[c]{@{}ccc@{}}
\toprule\addlinespace
\begin{minipage}[b]{0.15\columnwidth}\centering
parameter
\end{minipage} & \begin{minipage}[b]{0.18\columnwidth}\centering
lower.bound
\end{minipage} & \begin{minipage}[b]{0.18\columnwidth}\centering
upper.bound
\end{minipage}
\\\addlinespace
\midrule\endhead
\begin{minipage}[t]{0.15\columnwidth}\centering
$r$
\end{minipage} & \begin{minipage}[t]{0.18\columnwidth}\centering
1e-04
\end{minipage} & \begin{minipage}[t]{0.18\columnwidth}\centering
10
\end{minipage}
\\\addlinespace
\begin{minipage}[t]{0.15\columnwidth}\centering
$K$
\end{minipage} & \begin{minipage}[t]{0.18\columnwidth}\centering
1e-04
\end{minipage} & \begin{minipage}[t]{0.18\columnwidth}\centering
40
\end{minipage}
\\\addlinespace
\begin{minipage}[t]{0.15\columnwidth}\centering
$\theta$
\end{minipage} & \begin{minipage}[t]{0.18\columnwidth}\centering
1e-04
\end{minipage} & \begin{minipage}[t]{0.18\columnwidth}\centering
10
\end{minipage}
\\\addlinespace
\begin{minipage}[t]{0.15\columnwidth}\centering
$\sigma$
\end{minipage} & \begin{minipage}[t]{0.18\columnwidth}\centering
1e-06
\end{minipage} & \begin{minipage}[t]{0.18\columnwidth}\centering
100
\end{minipage}
\\\addlinespace
\bottomrule
\addlinespace
\caption{Parameterization range for the uniform priors in the Myers
model}
\end{longtable}

\newpage 
\newpage 

\subsection{Allen Model MCMC (Fig
S10-11)}\label{allen-model-mcmc-fig-s10-11}

\begin{figure}[htbp]
\centering
\includegraphics{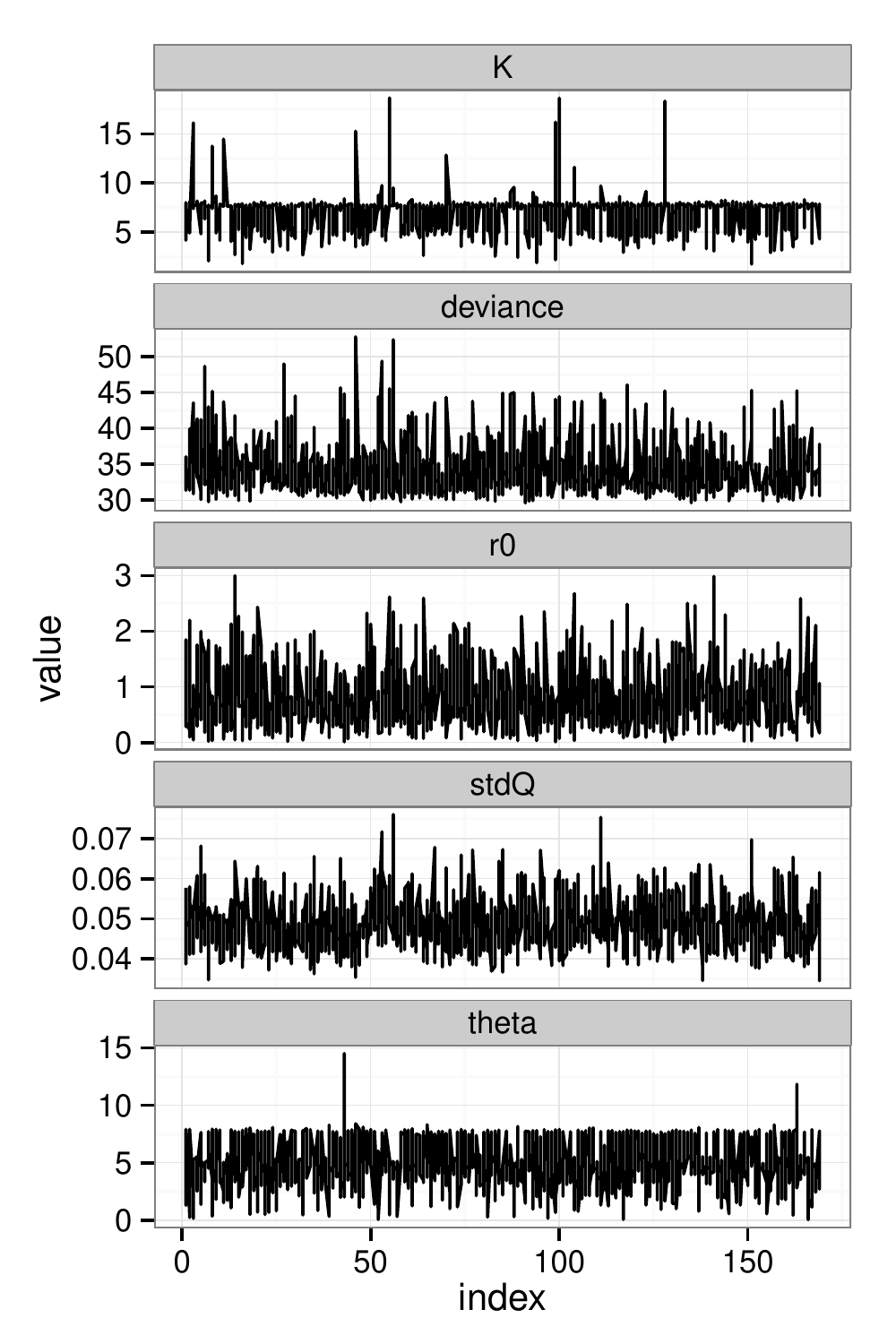}
\caption{Traces from the MCMC estimates of the Allen model show
reasonable mixing (no trend) and sampling rejection rate (no piecewise
jumps). stdQ refers to the estimate of $\sigma$; deviance is -2 times
the log likelihood.}
\end{figure}

\begin{figure}[htbp]
\centering
\includegraphics{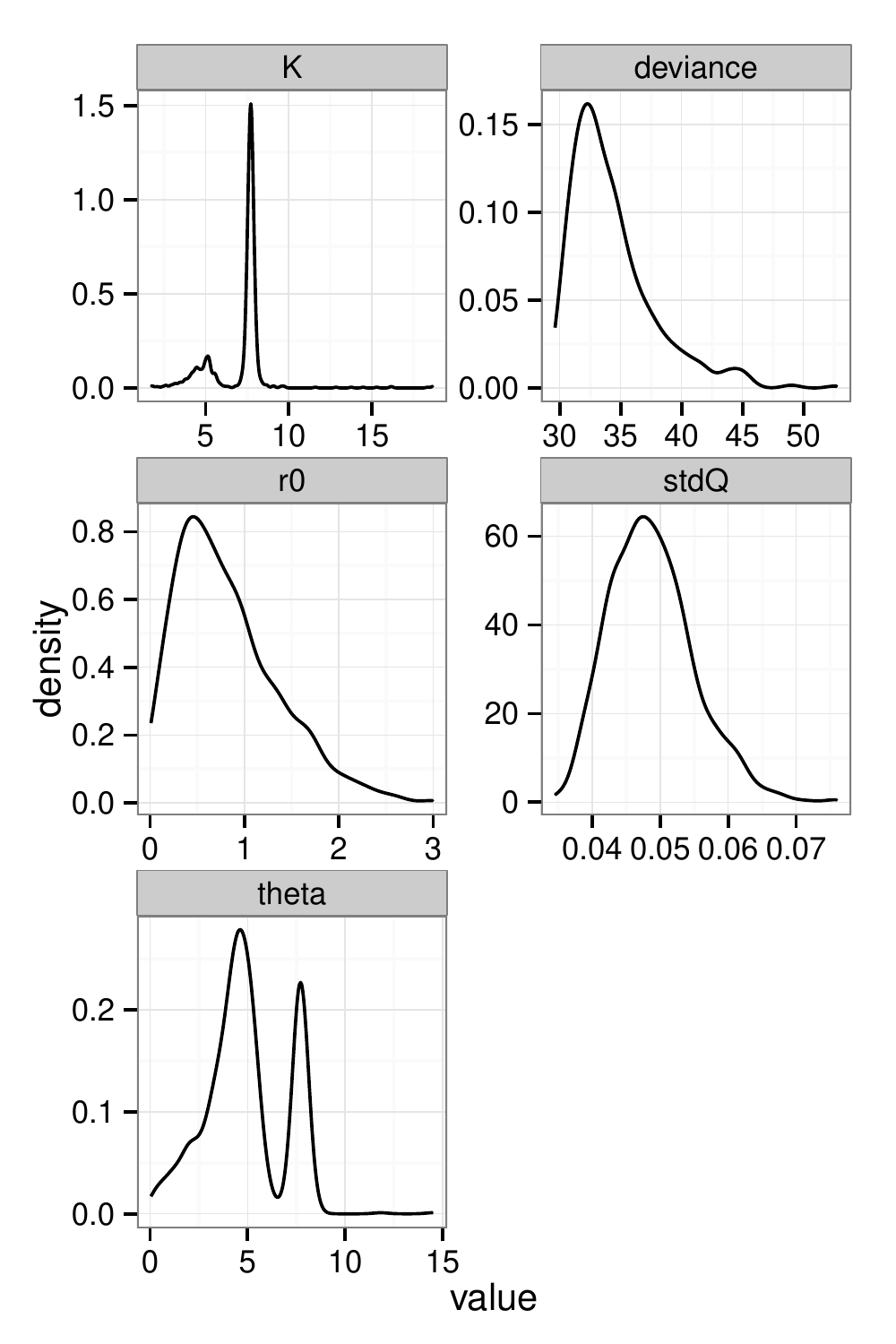}
\caption{Posteriors from the MCMC estimate of the Allen model. The Allen
model is the structurally correct model. Despite potential
identfiability issues in distinguishing between the stable and unstable
points ($K$ and $\theta$ respectively), the posterior estimates
successfully reflect both the the upper stable point ($K$), as well as
the significant probabilty of a tipping point ($\theta$) somewhere
between $K$ and extinction (0).}
\end{figure}

\begin{longtable}[c]{@{}ccc@{}}
\toprule\addlinespace
\begin{minipage}[b]{0.15\columnwidth}\centering
parameter
\end{minipage} & \begin{minipage}[b]{0.18\columnwidth}\centering
lower.bound
\end{minipage} & \begin{minipage}[b]{0.18\columnwidth}\centering
upper.bound
\end{minipage}
\\\addlinespace
\midrule\endhead
\begin{minipage}[t]{0.15\columnwidth}\centering
$r$
\end{minipage} & \begin{minipage}[t]{0.18\columnwidth}\centering
0.01
\end{minipage} & \begin{minipage}[t]{0.18\columnwidth}\centering
6
\end{minipage}
\\\addlinespace
\begin{minipage}[t]{0.15\columnwidth}\centering
$K$
\end{minipage} & \begin{minipage}[t]{0.18\columnwidth}\centering
0.01
\end{minipage} & \begin{minipage}[t]{0.18\columnwidth}\centering
20
\end{minipage}
\\\addlinespace
\begin{minipage}[t]{0.15\columnwidth}\centering
$X_C$
\end{minipage} & \begin{minipage}[t]{0.18\columnwidth}\centering
0.01
\end{minipage} & \begin{minipage}[t]{0.18\columnwidth}\centering
20
\end{minipage}
\\\addlinespace
\begin{minipage}[t]{0.15\columnwidth}\centering
$\sigma$
\end{minipage} & \begin{minipage}[t]{0.18\columnwidth}\centering
1e-06
\end{minipage} & \begin{minipage}[t]{0.18\columnwidth}\centering
100
\end{minipage}
\\\addlinespace
\bottomrule
\addlinespace
\caption{Parameterization range for the uniform priors in the Allen
model}
\end{longtable}

Gardiner, C. 2009. Stochastic Methods: A Handbook for the Natural and
Social Sciences (Springer Series in Synergetics). Page 447. Springer.

Mangel, M. 2006. The Theoretical Biologist's Toolbox: Quantitative
Methods for Ecology and Evolutionary Biology. Cambridge University
Press.

Xie, Y. 2013. Dynamic Documents with R and knitr. Chapman; Hall/CRCBoca
Raton, Florida.

\end{document}